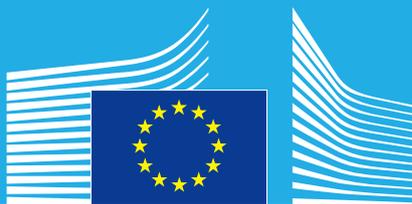

JRC TECHNICAL REPORTS

# Mobility and Economic Impact of COVID-19 Restrictions in Italy using Mobile Network Operator Data

A focus on the period November 2020 - February 2021

Vespe M., Minora U., Iacus S.M., Spyratos S., Sermi F., Fontana M., Ciuffo B., Christidis P.

2021

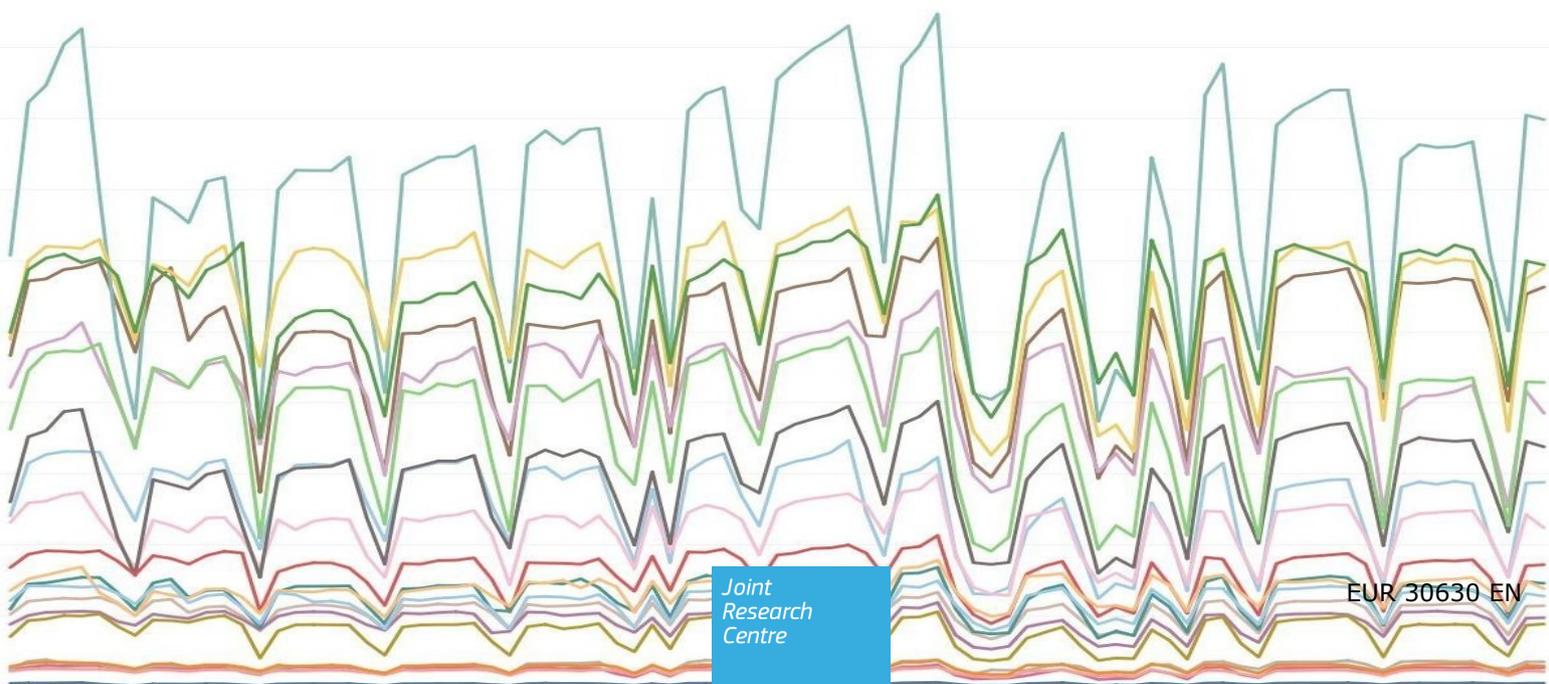

EUR 30630 EN

This publication is a Technical report by the Joint Research Centre (JRC), the European Commission's science and knowledge service. It aims to provide evidence-based scientific support to the European policymaking process. The scientific output expressed does not imply a policy position of the European Commission. Neither the European Commission nor any person acting on behalf of the Commission is responsible for the use that might be made of this publication.

Contact Information
Name: Michele Vespe
Address: Joint Research Centre, Via Enrico Fermi 2749, TP 26B, 21027 Ispra (VA), Italy
E-mail: michele.vespe@ec.europa.eu
Tel.: +39 0332 789154

EU Science Hub
https://ec.europa.eu/jrc

JRC124130

EUR 30630 EN

| PDF | ISBN 978-92-76-32080-7 | ISSN 1831-9424 | doi:10.2760/241286 |

Luxembourg: Publications Office of the European Union, 2021

# Contents

## Acknowledgements

The authors acknowledge the support of European MNOs (among which 3 Group - part of CK Hutchison, A1 Telekom Austria Group, Altice Portugal, Deutsche Telekom, Orange, Proximus, TIM Telecom Italia, Telefonica, Telenor, Tele2, Telia Company and Vodafone) in providing access to aggregate and anonymised data, an invaluable contribution to the initiative. The authors would also like to acknowledge the GSMA[1], colleagues from Eurostat[2] and ECDC[3] for their input in drafting the data request. Finally, the authors would also like to acknowledge the support from JRC colleagues, and in particular the E3 Unit, for setting up a secure environment to host and process of the data provided by MNOs, as well as the E6 Unit (the "Dynamic Data Hub" team) for their valuable support in setting up the data lake.

***Authors***
Michele Vespe, Umberto Minora, Stefano Maria Iacus, Spyridon Spyratos, Francesco Sermi, Matteo Fontana, Biagio Ciuffo, Panayotis Christidis.

---

[1]GSMA is the GSM Association of Mobile Network Operators.
[2]Eurostat is the Statistical Office of the European Union.
[3]ECDC: European Centre for Disease Prevention and Control. An agency of the European Union.

## Abstract

This work presents the analysis of the impact of restrictions on mobility in Italy, with a focus on the period from 6 November 2020 to 31 January 2021, when a three-tier system based on different levels of risk was adopted and applied at regional level to contrast the second wave of COVID-19. The impact is first evaluated on mobility using Mobile Network Operator anonymised and aggregate data shared in the framework of a Business-to-Government initiative with the European Commission. Mobility data, alongside additional information about electricity consuption, are then used to assess the impacts on an economic level of the three-tier system in different areas of the country.

## Highlights

— The varying levels of anti-coronavirus restrictions in different regions adopted in Italy has proved effective in systematically reducing mobility;

— The three-tier system of *red*, *orange* and *yellow* zones had a generalised impact on mobility at local, provincial and regional levels depending on the restrictions;

— Transitions to high-risk *red* zones resulted in a variation of total movements of -28% (-26.2% outward movements) and -36.9% (-40.5% outward movements) when downgrading from *orange* and *yellow* zones respectively;

— On average, at regional level, mobility in January 2021 has dropped by 29.9% with substantial geographic variation if compared to the one measured in January 2020, before any mobility restriction was applied in Italy;

— There's a significant geographic heterogeneity in the variation of mobility reduction, which is correlated to the pandemic incidence, suggesting an impact of the restrictions on mobility proportional to the epidemiological situation;

— Within the same level of restriction, after an initial period of transition, a systematic slight increase of mobility can be observed over time;

— An impact of the three-tier system on economic activity has been measured. The economic impact shows coherence with the direction of strengthening/weakening of the restrictions (e.g. a daily average variation of -0.33% for a *yellow* to *red* transition, vs +0.24% for a *red* to *orange*), and it correlates with the degree of such strengthening/weakening (e.g. a daily average variation -0.33% for *yellow* to *red* versus -0.15% for *yellow* to *orange*).

# 1  Introduction

Policy measures aimed at reducing the mobility of the general public had shown to be choice of governments to tackle the COVID-19 pandemic. They are proven to be generally effective (Zhou et al., 2020, Aleta and Moreno, 2020, Espinoza et al., 2020), and their ease of implementation renders them a very important measure for epidemic control. One of the most refined systems of mobility reduction in place is the Italian one: a ministerial decree was signed in Italy on 3 November 2020 containing stringent measures to slow COVID-19 infections[4]. The ministerial decree identified three areas at regional level (NUTS2[5]), corresponding to risk scenarios (*red*, *orange* and *yellow*) for which varying measures were envisaged. Regions are attributed different colours and restrictions depending on a number of epidemiological factors, and according to the level of pressure on hospitals and intensive care units. The classification is based on ordinances issued by the Italian Ministry of Health[6].

Without assessing the epidemiological benefits of these measures, the effects of these restrictions can be evaluated on mobility (see *e.g.* (Pepe et al., 2020)) and on the economy (Oliver et al., 2020). In this work, the assessment of such impacts is approached through the analysis of mobility insights extracted from aggregate and anonymised mobile positioning data collected in the framework of a Business-to-Government initiative between European Mobile Network Operators (MNOs) and the European Commission (see previous results (Santamaria et al., 2020), (Iacus et al., 2020a), (Iacus et al., 2021a), (Iacus et al., 2020b), (Iacus et al., 2021b) and (Christidis et al., 2021)). Differently from other mobility indicators publicly available such as the ones derived by social media, the data analysed in this work contain indications also on different components with respect to the geographic area under analysis, notably on internal, outward and inward movements. Moreover, depending on the provider, MNO data are characterised by higher frequency (up to hourly) and granularity (at municipality or district level), allowing a finer scale analysis with respect to openly available mobility indicators. These features make the data particularly suitable for the analysis of the impact of mobility restriction measures.

In the policy life-cycle, performing this type of analysis is extremely important in order to monitor the effectiveness of the measures introduced and perform adjustments in case they did not achieve the expected impacts. This ultimately helps understand what policies are most effective (Buckee et al., 2020) . For mobility restrictions to be effective in limiting and slowing down the diffusion of a virus, it is indeed essential that they are able to limit people movements at least beyond a certain level. Without quantitative evidence that this is achieved it is impossible to ensure the usefulness of the measure. In the end, mobility restrictions applied to citizens have several exemptions introduced to allow a sufficient level of service for the society and the fulfillment of the basic needs of the same citizens. These exemptions require time to citizens to be fully understood. This may induce a delay between the measure introduction and the achievement of its objectives and this delay poses challenges to the intention to limit the diffusion of the virus. Monitoring this delay can provide timely indications if additional measures need to be applied.

Another reason for implementing monitoring actions is to understand whether any measure applied may induce rebound effects that could jeopardise the effectiveness of the measure itself. As an example, imposing an earlier closure of commercial activities may induce an increase of mobility and the overcrowding of shops and other commercial areas before the new closure time, which in turn could even increase the diffusion of the virus.

Furthermore, another phenomenon that may occur and that monitoring mobility may help to keep under control is people relaxation to the measures. As already mentioned, people may need time to understand if and how to use the exemptions to mobility restrictions to fulfill their needs. If on the one hand this may delay the effectiveness of the measure, on the other hand it may also have the opposite effect. People may decide to initially restrict their mobility to beyond what is needed while relaxing it as soon as a better understanding of the exemptions is achieved (or as soon as it is understood that no suffi-cient enforcement mechanisms are in place). Therefore the initial successful application of the measure may be compromised by the subsequent relaxation period. Objective of the present report is to show simple ways in which the data collected in the aforementioned Business-to-Government initiative may provide extremely useful insights in the mobility dynamics induced by the various policy measures applied. The case study used in the analysis is Italy, with specific reference to one of the regions where the virus outbreak has had the strongest effects, the Lombardy Region.

---

[4]The Annex presents a list of legislative provisions that the Italian Government adopted between November 2020 and January 2021 to contain the spreading of COVID-19 pandemic and to manage the epidemiological emergency.

[5]NUTS: Nomenclature of Territorial Units for Statistics, Regulation (EC) 1059/2003 of the European Parliament and of the Council of 26 May 2003 on the establishment of a common classification of territorial units for statistics (NUTS).

[6]http://www.salute.gov.it/portale/nuovocoronavirus/dettaglioContenutiNuovoCoronavirus.jsp?id=5367&area=nuovoCoronavirus&menu=vuoto

The paper is structured as follows. Section 2 introduces the mobility data used to derive mobility insights. The measures adopted in Italy in the period 6 November 2020 - 31 January 2021 are described in Section 3. The impact of the relevant restrictions at regional (NUTS2), provincial (NUTS3) and city level mobility are analysed in Sec 4. The impact of the restrictions on the economy is then estimated using energy consumption data and proxies on mobility as discussed in Section 5. Finally, concluding remarks are given in Section 6.

## 2 Data

In this work, insights on mobility are extracted from Mobile Network Operator anonymised and aggregate data; epidemiological and economy and energy consumption data are also used.

### 2.1 Mobility data

Feeding epidemiological meta-population models, understanding the impact of confinement measures on mobility, as well as the role of mobility during the first phases of the epidemic were some of the uses identified in early April 2020 when the European Commission asked European Mobile Network Operators (MNOs) to share aggregate and anonymised mobile positioning data. Initiated by means of an exchange of letters, the terms of cooperation between MNOs and the European Commission are outlined by a Letter of Intent[7], which specifies that insights into mobility patterns of population groups extracted in the framework of this initiative are meant to serve the following purposes in the fight against COVID-19:

— "understand the spatial dynamics of the epidemics thanks to historical matrices of mobility national and international flows;

— quantify the impact of social distancing measures (travel limitations, non-essential activities closures, total lock-down,etc.) on mobility;

— feed SIR epidemiological models, contributing to the evaluation of the effects of social distancing measures on the reduction of the rate of virus spread in terms of reproduction number (expected number of secondary cases generated by one case);

— feed models to estimate the economic costs of the different interventions, as well as the impact of control extended measures on intra-EU cross border flows and traffic jams due to the epidemic; and

— cover all Member States in order to acquire insights relevant to Covid-19 to the entirety of the EU."

Data from European MNOs gradually began to be shared *pro-bono* with the European Commission covering 22 EU Member States plus Norway, on a daily basis, with an average latency of a few days, and in most cases covering historical data from January 2020 as in the case of Italy. Differently from mobility data derived *e.g.* from social media, MNO data can be processed to provide insights into human mobility at a level of granularity, timeliness, frequency of update, representativity (coverage of large fraction of the population) and transparency that make this data set uniquely positioned to continuously improve the response to an evolving emergency situation.

In compliance with the 'Guidelines on the use of location data and contact tracing tools in the context of the COVID-19 outbreak' by the European Data Protection Board (EDPB, 2020), MNO data shared in the framework of this initiative provide information of collective behaviours, and not on movements of individuals. Nonetheless, the data provide useful insights into human mobility patterns. The data is in the form of Origin-Destination Matrix (ODM) and describes the number of *'movements'* that have been recorded from the origin geographical reference area to the destination one over a specific time period. In general, an ODM is structured as a table reporting i) reference period (timestamp), ii) area of origin, iii) area of destination, iv) count of movements. Data quality aspects are deeply linked to privacy, data security and commercial confidentiality challenges as explained in (Vespe et al., 2021). *Mobility Indicators* are then produced by further space-temporal aggregating and normalising the ODMs provided by MNOs in order to ensure comparability across countries and operators. The indicators provide a daily time series of mobility according to the direction of the movements as internal (when it takes place within the same geographical area), inward (to a geographical areas), outward (from a geographical area), and total. More information about the *Mobility Indicators* and their applications can be found in (Santamaria et al., 2020). The 'Staying safe from COVID-19 during winter' strategy adopted by the European Commission in December 2020 mentions: "Insights into mobility patterns and role in both the disease spread and containment should ideally feed into such targeted measures. The Commission has used anonymised and aggregated mobile network operators' data to derive mobility insights and build tools to inform better targeted measures, in a Mobility Visualisation Platform, available to the Member States. Mobility insights are also useful in monitoring the effectiveness of measures once imposed." (COM(2020), 786, 2020). The products are currently expanded to feed early warning mechanisms to detect anomalies in usual mobility patterns such as gatherings (De Groeve et al., 2020) and (Iacus et al., 2021b).

---

[7]European Commission and GSMA partners on Data4Covid - https://www.gsma.com/gsmaeurope/resources/d4c/

## 2.2 Additional Data

The present work makes also use of epidemiological data on regional number of cumulative COVID-19 cases from the Italian Department of Civil Protection normalised to the population provided by the Italian National Institute of Statistics (ISTAT) as described in Section 4.3. The impact on the economy is estimated using World Bank data for national GDP[8], Terna Rete Italia for monthly energy demand[9] and Eurostat for NUTS2-level GDP data[10], as described in Section 5.

## 3 Mobility containment measures in Italy from 6 November 2020 to 31 January 2021: the "three-tier" system

A ministerial decree signed on 3 November 2020 in Italy contained stringent measures to counter the spread of COVID-19[11]. The decree identifies three main areas, corresponding to three risk scenarios (see Figure 19 in the Annex), for which different measures are envisaged: most restrictive (*red*), slight restrictive (*orange*) and remaining regions (*yellow*). In addition, the measures included a night curfew from 22:00 to 5:00, the closure of swimming pools, gyms, cinemas, theatres and museums, whereas shopping centres were closed throughout the country at weekends. Transports were also limited to 50% capacity.

Entering, leaving, or moving within towns or cities was prohibited within **red zones**. Movements were authorised for reasons of necessity, such as work, taking children to school (distance learning applies to the final two years of middle school and high school), or health. Coffee places, restaurants and most shops were closed, whereas supermarkets, and pharmacies were allowed to remain open.

Coffee places and restaurants were closed also in **orange zones**, whereas shops were open. In addition, distance learning applies to high-schools and movements were allowed within towns or cities, but not between them. It was therefore not allowed any movement across regions.

In **yellow zones**, restaurants and bars were open until 18:00, and shops were open. Shopping centres were closed during weekends. Movements were allowed within the region. Movements from one *yellow* region to another were allowed until 1 February 2021, when these movements were then forbidden. Figure 1 shows the time evolution of the restrictions for all regions in Italy.

---

[8]https://data.worldbank.org/indicator/NY.GDP.MKTP.CD?locations=IT
[9]https://www.terna.it/it/sistema-elettrico/pubblicazioni/rapporto-mesile
[10]https://ec.europa.eu/eurostat/web/products-datasets/-/tgs00004
[11]The Annex presents a list of legislative provisions that the Italian Government adopted between November 2020and January 2021 to contain the spreading of COVID-19 pandemic and to manage the epidemiological emergency.

During the Christmas period and close to the new year, from 24 December 2020 to 10 January 2021, a time-series of transitions was planned uniformly throughout Italy, and the same restrictions applied in all regions.

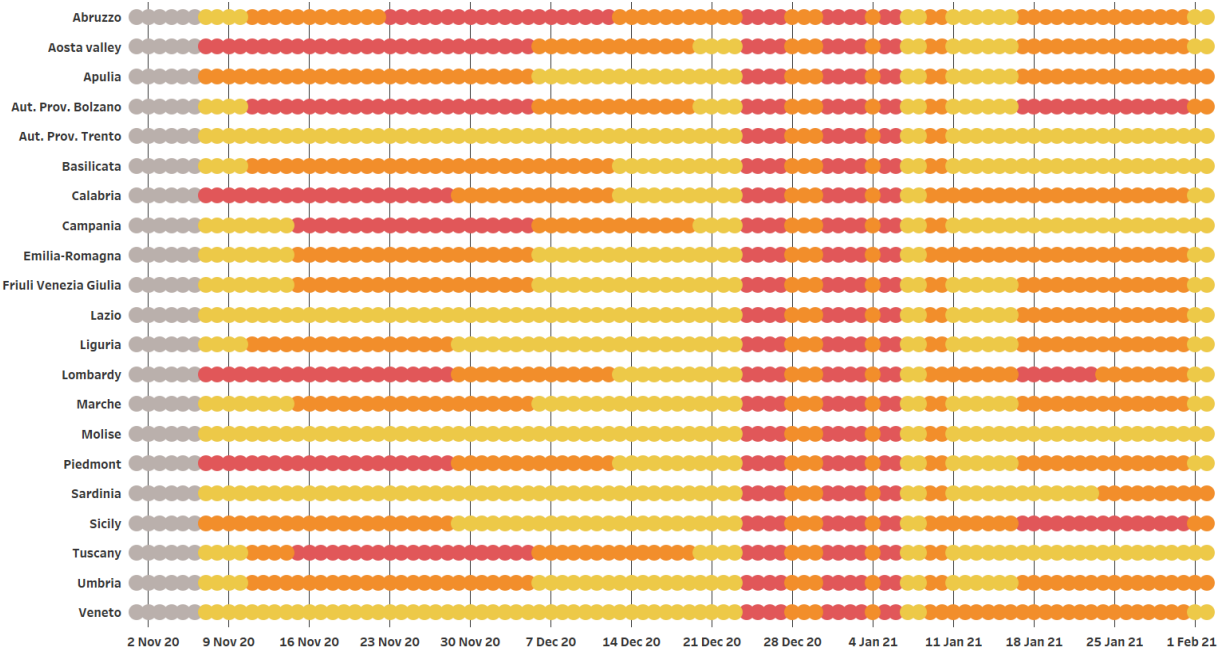

Figure 1: Time series of mobility measures taken in all Italian Regions by level of restriction: *grey* (before the ministerial decree on 6 November), *red* (high), *orange* (medium) and *yellow* (low).

On 1 February 2021 Italy eased the restrictions in most regions. With the exception of only five *orange* regions, all the others turned *yellow*. However, it is worth noting that from the same day onward, unlike the measures taken previously, the movements between *yellow* regions were prohibited.

## 4  Impact of restrictions on mobility

Mobile positioning data are used to quantify the impact on human mobility of regional and national containment measures adopted to reduce the spreading of the virus. Subsection 4.1 investigates the effect of curfews measures alone, adopted by regional authorities ahead of the ministerial decree of 6 November 2020. Finally, Subsection 4.2 shows the analysis of mobility variations in correspondence of the more structured set of restrictions on mobility, which also include curfew measures, of the national three-tier system at different geographical scales, from the aggregate regional to the city levels.

### 4.1  Impact of regional curfew measures from 22 October to 5 November 2020

Before the ministerial decree of 6 November 2020 (described in Section 3), regional authorities together with the Italian Ministry of Health imposed a night curfew in Lombardy[12]. The curfew, valid from 23:00 to 5:00 as of Thursday 22 October 2020, was accompanied by a set of additional measures calling large shopping centres to be closed at weekends. The impact of the curfew on human mobility can be clearly observed for Milan province in Figure 2, which shows a marked drop of overall daily movements (outward + inward movements within the time window 23:00-24:00) ranging from about one third on weekdays to over two thirds on Fridays and Saturdays. It is worth recalling that, because of the typical weekly mobility patterns, e.g. showing internal mobility minima on Sundays ((Santamaria et al., 2020)), in the attempt to extract mobility trends over different weeks, a given day should be always compared with the same weekday. This is why in the figure days are grouped by weekdays, in order to ease their comparison in time. The first three bars for each weekday are relative to days preceding the entry into force of the curfew. The mobility reduction in the night hours due to the curfew appears generally more pronounced for Milan province.

---
[12]Decree of the Ministry of Health together with Lombardy Region of 21 October 2020: Further prevention and management measures for the COVID-19 epidemiological emergency. ORDINANZA AI SENSI DELL'ART. 32 DELLA LEGGE 23 DICEMBRE 1978, N. 833 IN MATERIA DI IGIENE E SANITÀ PUBBLICA E DELL'ART. 2, COMMA 2 DEL DECRETO-LEGGE 25 MARZO 2020, N. 19 CONVERTITO CON MODIFICAZIONI DALLA L. 22 MAGGIO 2020, N. 35 https://www.gazzettaufficiale.it/eli/id/2020/10/21/20A05821/sg

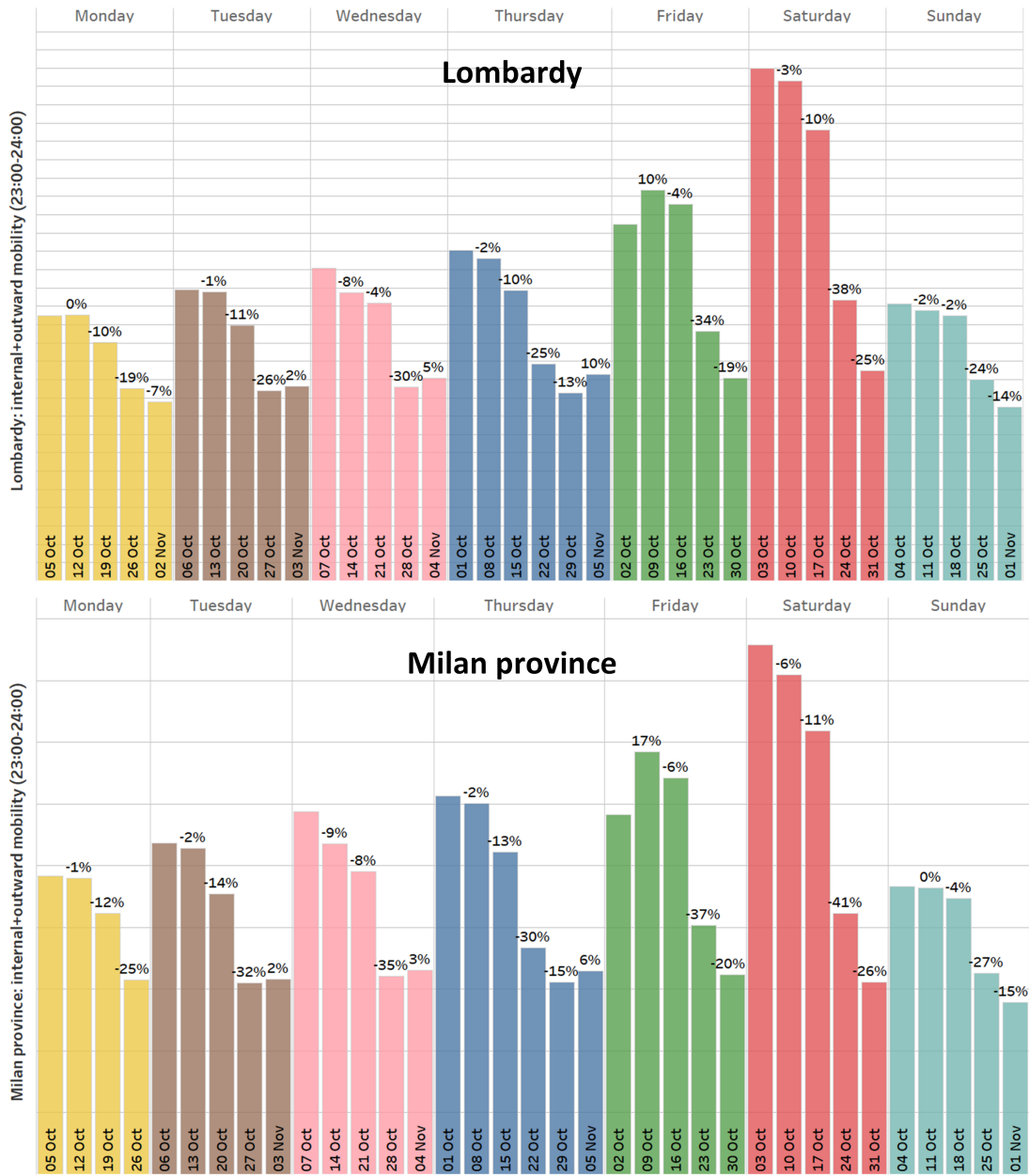

Figure 2: Daily outward and internal mobility (between 23:00 and 24:00) for Lombardy (top) and Milan province (bottom). The reference period is Thursday 1 October - Thursday 5 November 2020. On Thursday 22 October, a regional decree entered into force, imposing curfew between 23:00 and 5:00. Percentages show the relative variation of mobility respect to the same weekday of the previous week. Mobility data for 2 November is missing. The y-axes of the two charts are not in scale with each other.

The chart in Figure 3 is an alternative way to visualise the impact of the night curfew imposed by the regional authorities on mobility in Lombardy (left) and in Milan province (right). The week before (purple) and that after (orange) the entry into force of the curfew are compared.

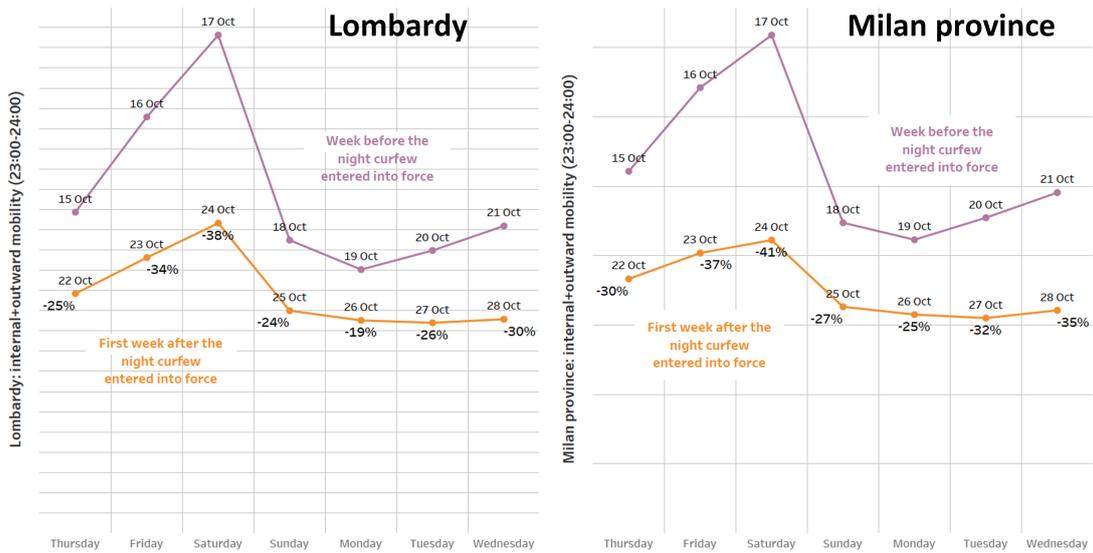

Figure 3: Daily outward and internal mobility (between 23:00 and 24:00) for Lombardy (left) and Milan province (right). The seven days before the entry into force of the night curfew (purple) are compared with the first seven days after the entry into force (orange). Percentages show the relative variation of mobility respect to the same weekday of the previous week. The y-axes of the two charts are not in scale with each other.

The reduction of mobility in the eleven other provinces in Lombardy was analogous to that of Milan province, but with a slightly different effect. Figure 3 shows the daily internal mobility (between 23:00 and 24:00) in Lombardy from 1 October 2020 to 1 February 2021. The three-tier system entered into force on 6 November extending the night curfew measure from 22:00 to 5:00, corresponded to an initial further decrease of mobility (internal + outward) in Lombardy between 23:00 and 24:00. The effect of the additional restrictions to mobility, adopted at National level to limits the spreading of the virus around Christmas and the new year's eve, are evident in the figure.

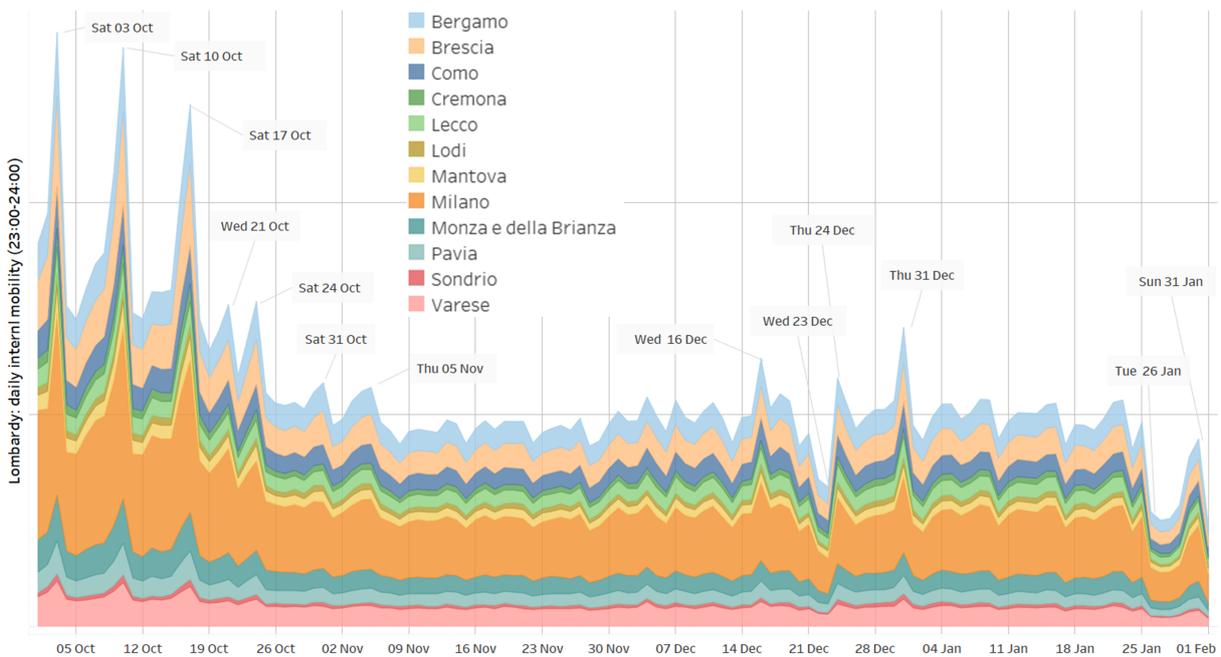

Figure 4: Daily internal mobility (between 23:00 and 24:00) in Lombardy from 1 October 2020 to 1 February 2021. Internal mobility is disaggregated by province. Labelled days in x-axis are Mondays. The most pronounced weekly peaks of mobility are indicated by additional labels.

## 4.2 Impact of national restrictions from 6 November 2020 to 31 January 2021

The impact of the structured set of restrictions on mobility introduced by the national three-tier system (Section 3) is investigated at different geographical scales, from the aggregate regional to the city levels.

### 4.2.1 Regional level

As described in the Data section, the mobility indicators provide a daily time series of the number of aggregate and anoymised movements of mobile phone users between geographical areas (see also (Santamaria et al., 2020)). In the following analysis, all indicators are normalised using a reference value that is common to all granularity levels and mobility components (*i.e.* internal, inward and outward). For example, indicators at NUTS2 or NUTS3 level of internal, inward or outward mobility are all normalised to the same reference number in all regions or provinces in Italy, allowing for geographical comparison of mobility across different mobility components. Figures 5, 6, 7, and 8, show the total, internal, inward, and outward daily mobility trends (respectively) for each region in Italy between 1 November 2020 and 1 February 2021. The same figures also show the succession of the colour-coded restrictions in each region, which allows to visually relate possible changes in the mobility trend during the transition from one level of restriction to another (see also Table 1).

The impact of the different restrictions, the reaction time after the transition, as well as a generalised progressive increase of mobility within the same colour scheme are clearly visible in almost all the regions for all the movements components, either intra-regional and between regions. The link between the type of transition (*e.g.* from *yellow* to *red* or from *orange* to *yellow*) and the extent of variation of mobility can also be observed, demonstrating a clear effectiveness of the restrictions in reducing mobility. In the following part, such effect is further analysed quantified.

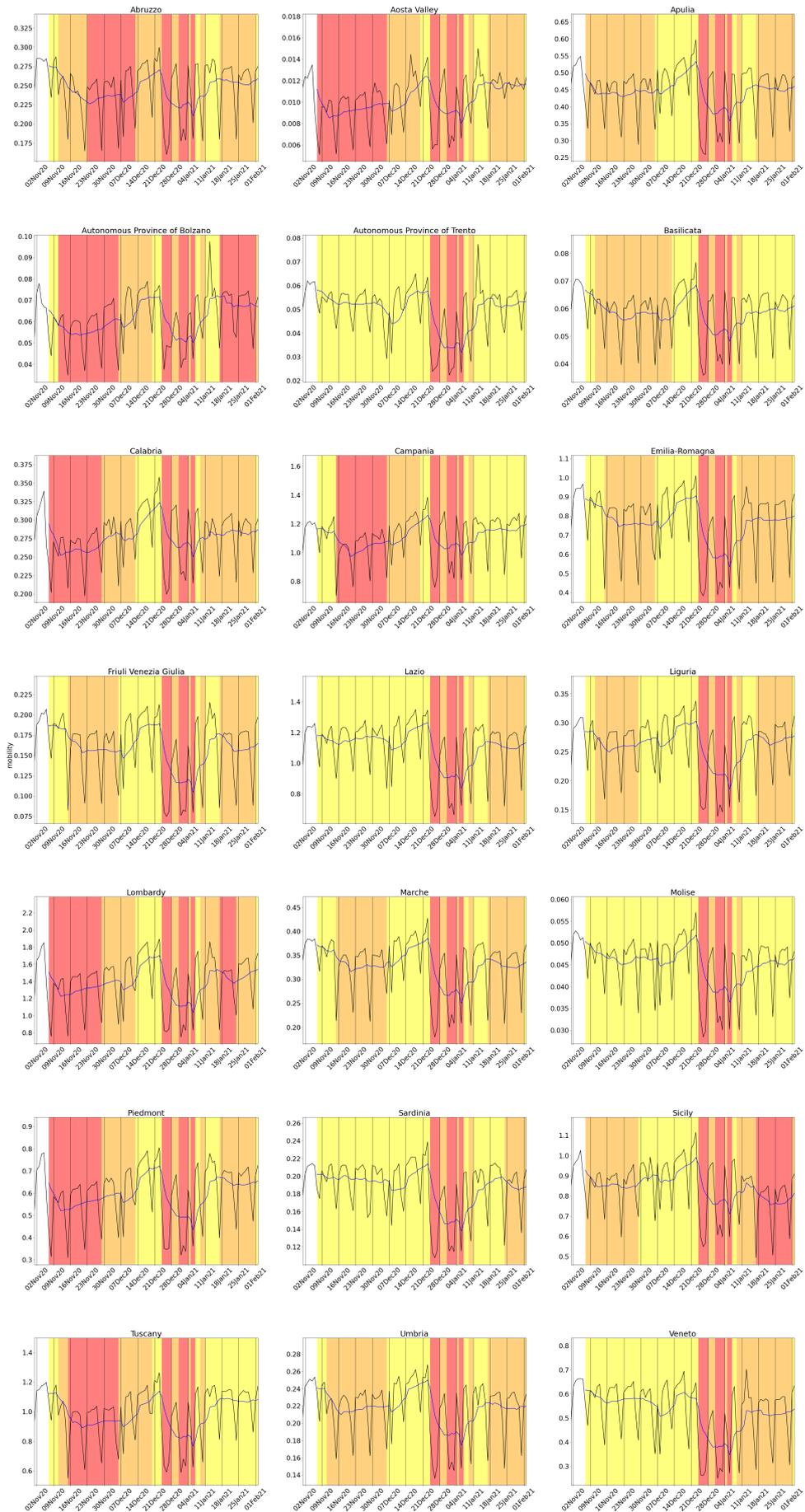

Figure 5: Time evolution of colour-coded restrictions in different regions in Italy, their evolution over time, *total mobility* (black line) and weekly average mobility (blue line).

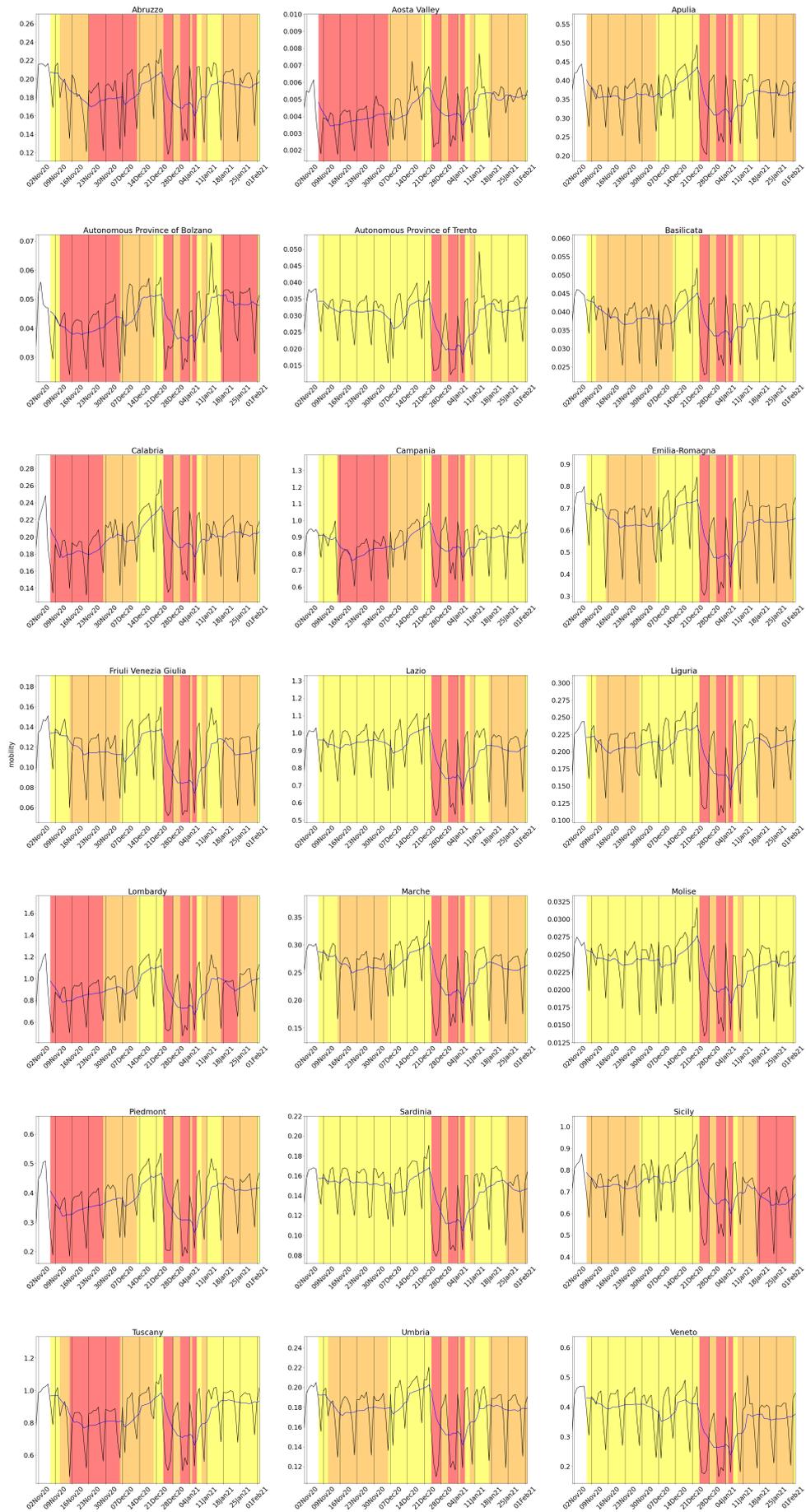

Figure 6: Time evolution of colour-coded restrictions in different regions in Italy, their evolution over time, *internal mobility* (black line) and weekly average mobility (blue line).

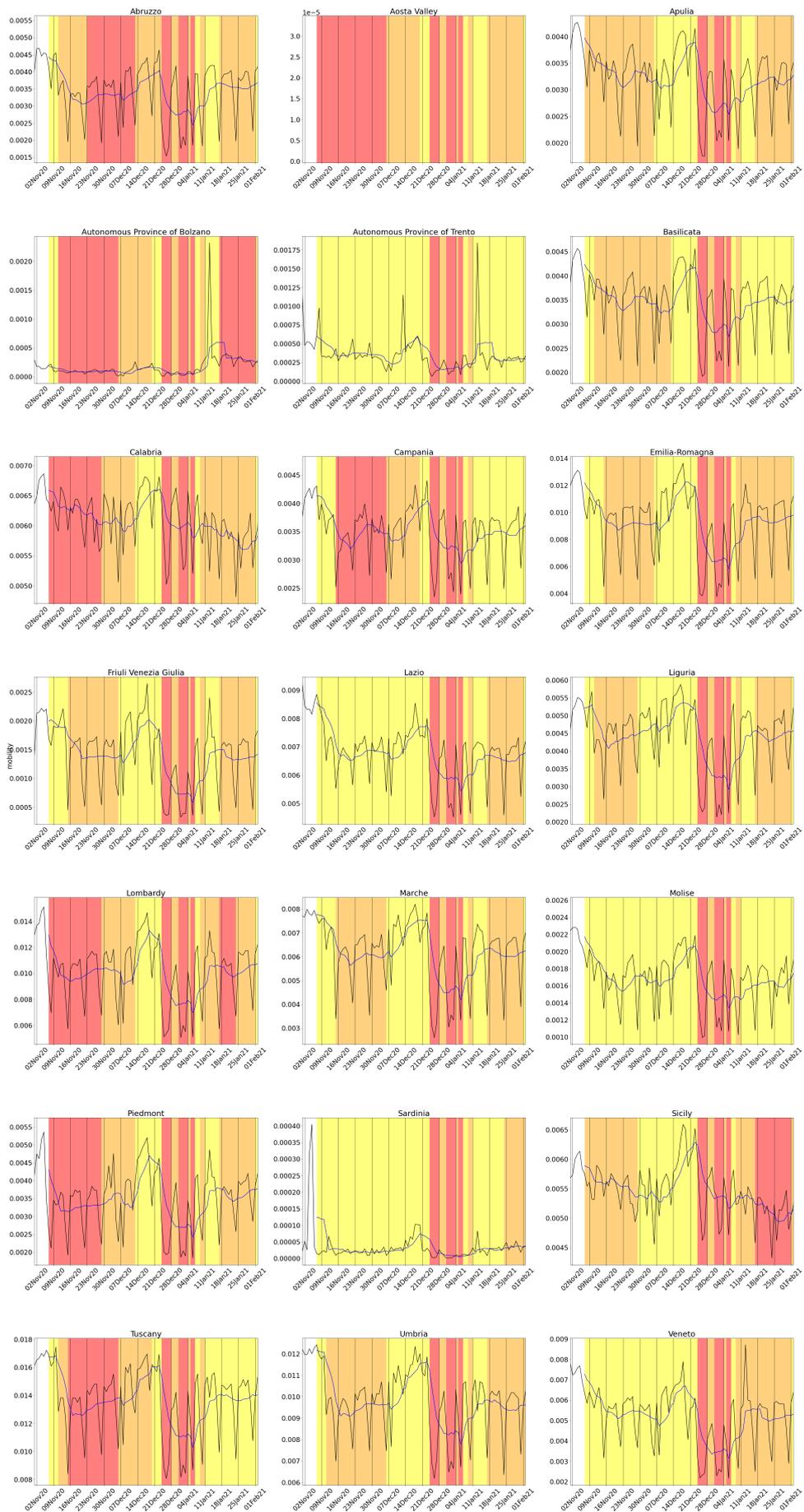

Figure 7: Time evolution of colour-coded restrictions in different regions in Italy, their evolution over time, *inward mobility* (black line) and weekly average mobility (blue line).

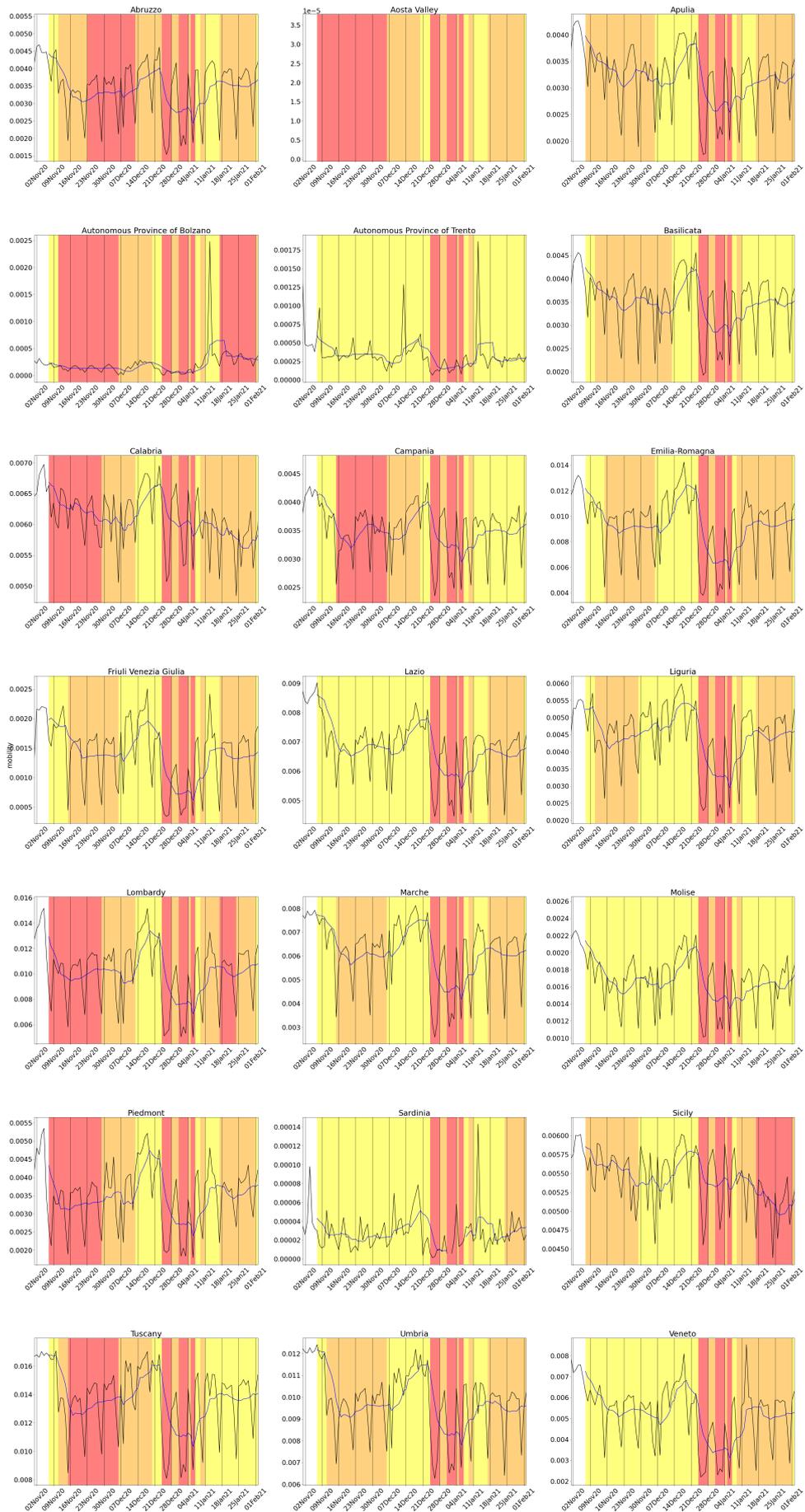

Figure 8: Time evolution of colour-coded restrictions in different regions in Italy, their evolution over time, *outward mobility* (black line) and weekly average mobility (blue line).

The impact of the different transitions of the three-tier system on mobility is now analysed over time and across regions in Italy. Such impact is computed by calculating, for each transition between colors of a region, and for each mobility component, *i.e.* internal, outward, inward and total.

The first step is to compute, for each region $i$ in the set of italian regions $I$, for each indicator $j$ in the set of indicators that identify the type of movement (internal, inward, outward or total), for each type of transition $k \in K$ is the type of transition index (*e.g. orange* to *red*, *red* to *yellow* etc.) and for each replicate of the transition of type $k$, $l \in L_k$, taking into account that the same transition can happen multiple times for the same region and type of movement, a set of mobility indices pre- and post- the transition, indicated as $\tilde{\mathcal{I}}_{ijkl}^{pre}$ and $\tilde{\mathcal{I}}_{ijkl}^{post}$ as the arithmetic mean of the indicator $k$ respectively before and after the transition uniquely identified by $i, j, l$.

Having these average mobility indices, several possible metrics can be calculated to describe the impact of the restrictions: in Table 1, the average percentage variation for each transition type, and for each index, is computed as follows:

$$\frac{\frac{\tilde{\mathcal{I}}_{\cdot jk\cdot}^{post}}{|I||L_k|} - \frac{\tilde{\mathcal{I}}_{\cdot jk\cdot}^{pre}}{|I||L_k|}}{\frac{\tilde{\mathcal{I}}_{\cdot jk\cdot}^{post}}{|I||L_k|}} * 100$$

where $\tilde{\mathcal{I}}_{\cdot jk\cdot}^{post}$ is the summation of $\tilde{\mathcal{I}}_{ijkj}^{post}$ over the different $i$ and $j$, $L_1$ and $L_2$ are the set of days to and $|\cdot|$ indicates the cardinality of the set.

| Transition | Internal | Inward | Outward | Total |
|---|---|---|---|---|
| yellow → orange | -18.5% | -19.9% | -19.9% | -18.5% |
| yellow → red | -36.8% | -40.5% | -40.5% | -36.9% |
| orange → red | -28.1% | -26.3% | -26.2% | -28.0% |
| red → orange | +43.1% | +38.7% | +38.6% | +43.0% |
| orange → yellow | +24.7% | +26.1% | +25.9% | +24.7% |

Table 1: Overall variation of internal, outward and inward mobility depending on the transitions.

Table 1 is an indication of the overall impact of the transitions on the different type of movement. It is worth noting that, since the volume of internal movements in the region is substantially greater than the inward and outward ones, the impact of the restrictions on total movements is biased towards the internal ones. It is also important to underline that the transition *yellow* to *orange* seems having the lowest impact in relative terms, but it refers to starting conditions of highest absolute volumes of movements as in *yellow* risk (see Figure 5).

In Figure 9 calculations are performed using the following formula:

$$\frac{\frac{\tilde{\mathcal{I}}_{ijk\cdot}^{post}}{|L_k|} - \frac{\tilde{\mathcal{I}}_{ijk\cdot}^{pre}}{|L_k|}}{\frac{\tilde{\mathcal{I}}_{ijk\cdot}^{post}}{|L_k|}} * 100$$

That is the average percentage variation for each transition type, each index, and each region. It can be seen that a specific transition has a generalised and aligned variation of mobility, although this is not uniform, pointing to geographic differences. This aspect is further analysed in Section 4.3.

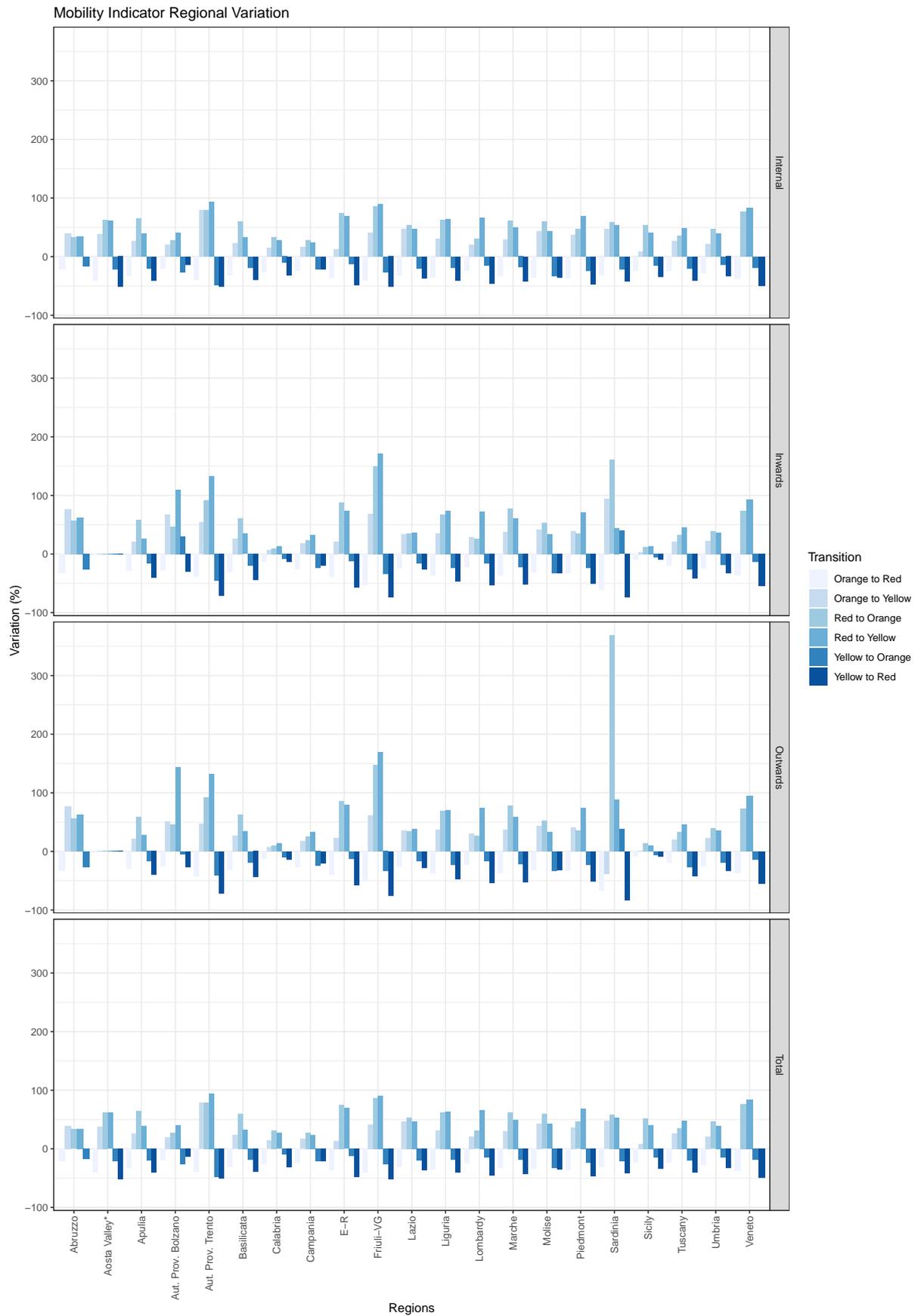

Figure 9: Barplot of percent regional variations of mobility indicators: internal, inward, outward and total (top to bottom).

By analysing the mobility data, it is clear that the transition into a new restriction regime in most cases leads to a rapid change in the mobility indicators. When the new phase is stricter (e.g. from *yellow* to *orange*, or from *orange* to *red*) mobility falls abruptly, while if the new phase is more relaxed (*e.g.* from *red* to *orange*, or from *orange* to *yellow*) mobility returns to higher level. Once in a new phase, mobility generally rises with time towards the levels registered before the new restrictions are enforced. A possible interpretation could be that the population is quick to adapt to the new restriction situation, but tend to either comply less or understand better the perimeters and margins of movements allowed by such restrictions as time passes. A regression analysis at province level (NUTS3) is here applied to Lombardy mobility indicators shown in Figure 10 to quantify the relaxation effect during this phase, see Table 2.

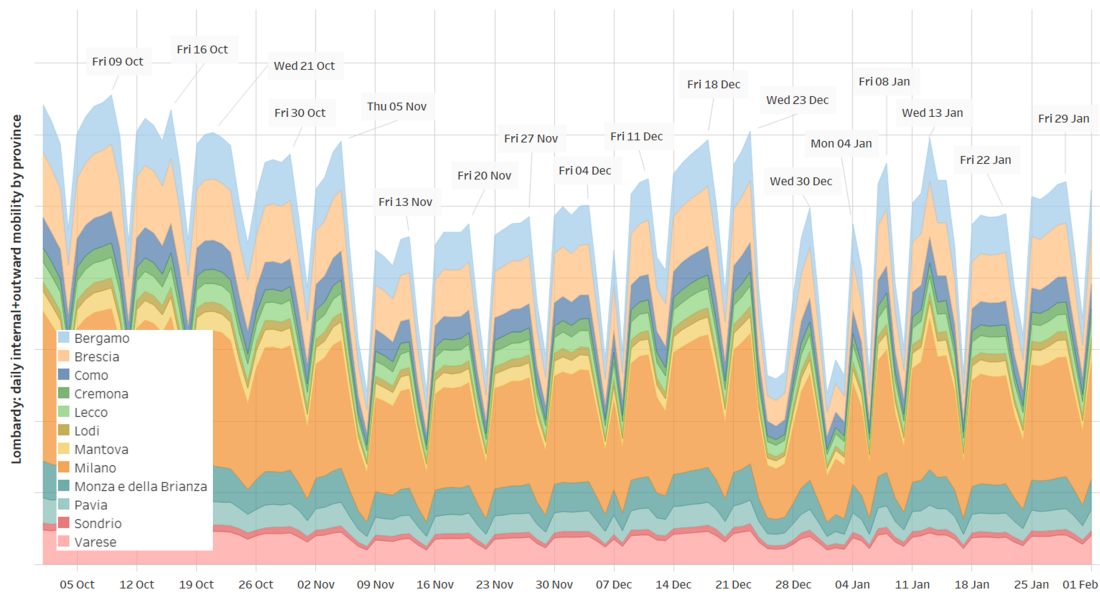

Figure 10: Daily internal mobility by province in Lombardy (Oct. 2020 to Jan. 2021). Labelled days in x-axis are Mondays. The weekly peak of mobility is indicated by additional labels.

|  | Estimate | Std. Error | t value | $Pr(>|t|)$ |
|---|---|---|---|---|
| (Intercept) | 0.06932 | 0.01196 | 5.794 | 9.29e-09*** |
| Orange→Red | -0.08713 | 0.02266 | -3.845 | 0.000128*** |
| Orange→Yellow | 0.25205 | 0.02620 | 9.620 | < 2e-16*** |
| Red→Orange | 0.14379 | 0.01742 | 8.257 | 4.91e-16*** |
| Yellow→Orange | -0.28374 | 0.03852 | -7.365 | 3.80e-13*** |
| Yellow→Red | -0.33817 | 0.02265 | -14.932 | < 2e-16*** |
| weekend/bank holiday | -0.10483 | 0.01446 | -7.249 | 8.64e-13*** |

Signif. codes: 0 '***' 0.001 '**' 0.01 '*' 0.05 '.' 0.1 ' ' 1
Residual standard error: 0.2104 on 961 degrees of freedom
Multiple R-squared: 0.4075, Adjusted R-squared: 0.4038
F-statistic: 110.1 on 6 and 961 DF, p-value: < 2.2e-16

Table 2: Regression Analysis of the relaxation effect after the transition to stricter restrictions. The dependent variable is the percentage change in mobility compared to the previous week.

The dependent variable of the regression model presented in Table 2, is the percentage change in mobility compared to the week before. The regression model takes into account the zone colour-code (i.e. level of restriction) on current day and on the same day of the previous week, as well as whether it is a weekend or bank holiday. The intercept means that the underlying trend is positive, and it is translated to an approximate 7% increase on mobility per week while the restrictions remain the same. In other word this means that people tend to move 1% more per day and this is referred to as the relaxation to the restrictions.

The trend of mobility reduction and the subsequent increase during the first lockdown in Italy in early March 2020 was different if compared to the one observed in the second. As shown in Figure 11 the drop of mobility was more pronounced in March, as a consequence of the the stricter measures in place compared to the ones implemented during the three-tier system. In addition, in March 2020, the same restrictions applied to the whole country with no regional variations. Compared to the first week (9-16 March) of the first lockdown, the mobility decreased by 24% in second week (16-22 March), and by a further 11% in the third week (23-29 March) when it reached its lowest point. From that week (23-29 March) and onward, mobility started gradually recovering by 3% in the fourth week followed by another 6% in the fifth week.

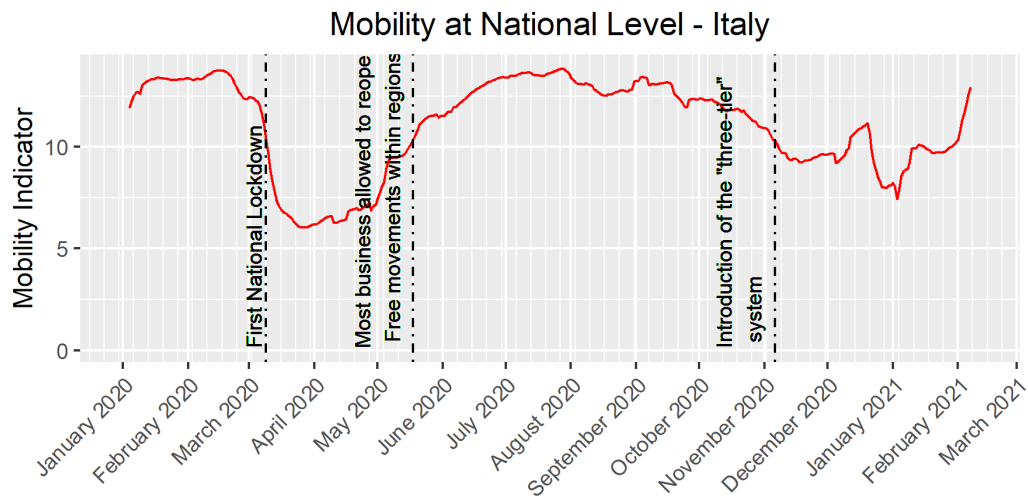

Figure 11: Total mobility aggregated at national level during the first lockdown in March 2020 and the three-tier system from November 2020 to 31 January 2021.

The effects of the ministerial decree setting up the three-tier system are now analysed in detail for the Lombardy Region, which became a *red* area. After three weeks, on Sunday 29 November, Lombardy transitioned from *red* to *orange*.
Figure 12 shows the daily mobility for the region and its relative variation respect to the same weekday of the previous week.

The highest relative reductions of mobility appear to be related with festivity:

— -56% (*yellow→red*): Friday 18 Dec - Friday 25 Dec (Christmas);

— -49% (*yellow→red*): Saturday 19 Dec - Saturday 26 Dec (St. Stephen's Day);

— -47% (*orange→red*): Wednesday 30 Dec - Wednesday 6 Jan (Epiphany);

— -41% (*orange→orange*): Tuesday 1 Dec - Tuesday 8 Dec (Feast of the Immaculate Conception).

Only the following highest reduction is not related to a festivity, but to the first Sunday since the entry into force of the decree:

— -38% (*before 3-tier→orange*): Sunday 1 Nov - Sunday 8 Nov.

Also the highest relative increases are related to festivity:

— +134% (*red→yellow*): Friday 1 Jan (New Year's Day) - Friday 8 Jan;

— +115% (*red→orange*): Wednesday 6 Jan (Epiphany) - Wednesday 13 Jan;

— +89% (*orange→yellow*): Tuesday 8 Dec (Feast of the Immaculate Conception) - Tuesday 15 Dec;

The following highest relative increase of mobility correspond to the first *yellow* day after 36 days from the entry into force of the decree:

— +44% (*orange→yellow*): Sunday 6 Dec - Sunday 13 Dec.

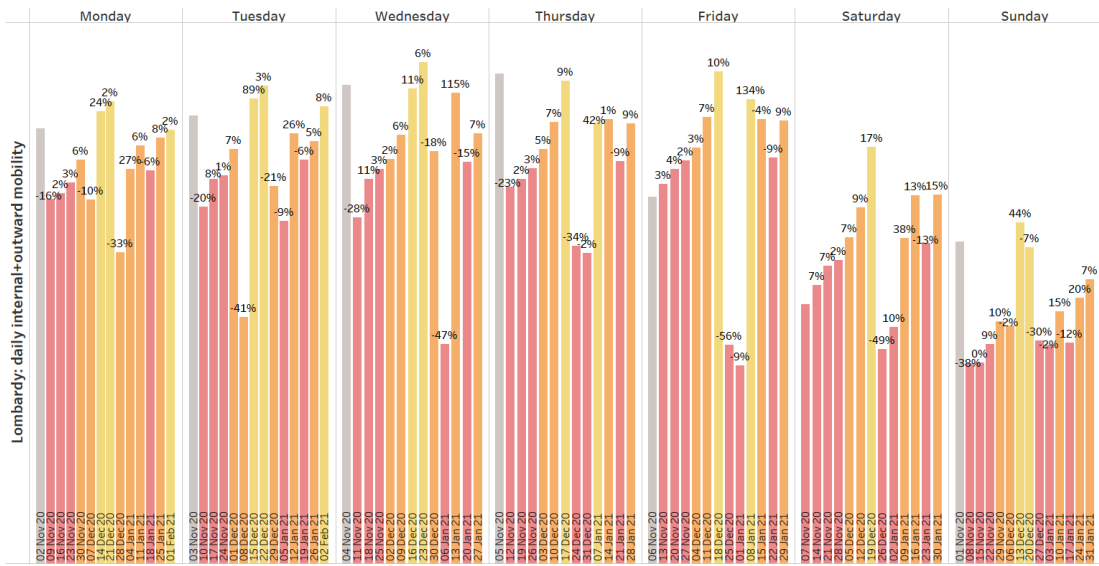

Figure 12: Total daily mobility for Lombardy (bar height) by weekdays and its relative variation respect to the same day of the previous week. Colours are associated to the three-tier system; grey bars are relative to days before the entry into force of the decree.

As a general consideration, which holds also for the other Italian regions, the evident decrease of mobility due to the transition to an higher risk-level tier (*i.e.* from *yellow* to *orange* or from *orange* to *red*), is mitigated throughout the same tier highlighting the relaxation effect on restrictions as discussed above.

### 4.2.2 Province level

The three-tier scheme was effective nationally as of 6 November 2020. It is important to understand to what extent such measures are adopted uniformly at province level.

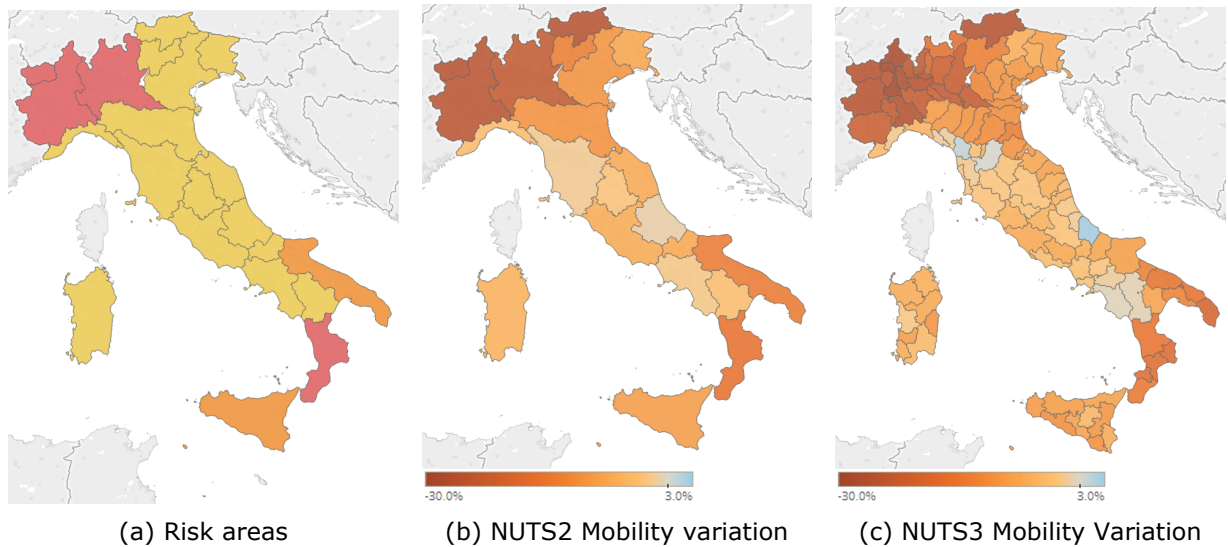

(a) Risk areas  (b) NUTS2 Mobility variation  (c) NUTS3 Mobility Variation

Figure 13: Distribution of colour-coded restrictions in Italy following the ministerial decree of 6 November 2020 (a) and variation of total mobility between Monday 27 October 2020 and Monday 10 November 2020 at regional (b) and provincial level (c).

In Figure 13a, the map shows the colours of the regions across the country. As also illustrated by the first column of Figure 1, on 6 November four regions were set as *red* zones (Calabria, Lombardy, Piedmont and Aosta Valley), two *orange* (Apulia and Sicily) and the remainder *yellow*. The effect of the restrictions can be seen in Figure 13b, reporting the variation of total mobility between Tuesday 27 October and Tuesday 10 November. The relative drop of mobility is more pronounced for *red* zones, notably -26.1% in Aosta Valley,

-25.3% in Piedmont and -23.5% in Lombardy where the largest reduction of number of movements was registered in Italy at the turn of the entry onto force of the ministerial decree. In Figure 13c, the relative variation of total movements in the same time period is shown at NUTS3 level. Provinces are consistently aligned with the measures taken at the level of their region.

### 4.2.3 Local level

The impact of the restrictions on mobility enforced at national level with different levels according to the three-tier systems of regional zones is now analysed at more local level. Between Tuesday 27 October and Tuesday 10 November 2020, at the turn of the entry onto force of the ministerial decree identifying Lombardy as a high risk area and therefore enacting the most stringent measures, data shows an average -23.5% reduction of total mobility, notably -26% for the province of Milan.

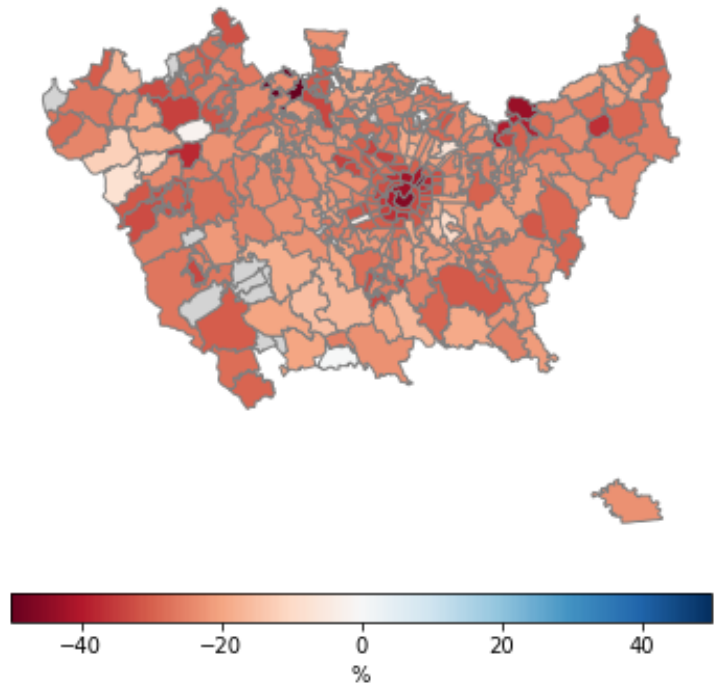

Figure 14: Variation of total mobility between 27 October and 10 November 2020 in the province of Milan.

Figure 14 illustrates a map of total mobility reduction over different census tracts in the province of Milan. The distribution of the mobility variation in Figure 18 shows an overall adherence to the restrictions adopted at regional scale, with a drop of up to -40% in high population density areas as the city of Milan located in the center of the map, or in proximity of large commercial centres or shopping areas (*e.g.* North-East and North-West of Milan, in the areas of Lainate and Carugate).

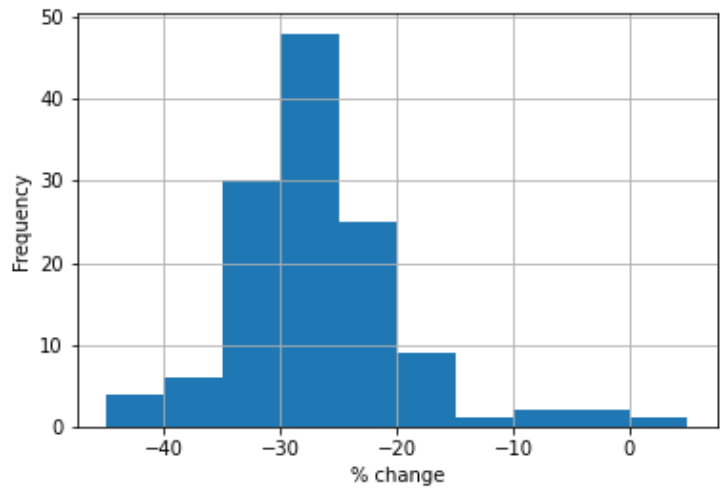

Figure 15: Distribution of total mobility variation between 27 October and 10 November 2020 of census tracts in the province of Milan.

## 4.3 Mobility comparison between January 2020 and January 2021

A general decrease in total mobility is observed comparing January 2021 against January 2021 (Figure 16). The mobility in January 2021 during working days (which are usually when most movement of the week is registered), is in general at the same level of the 2020 mobility during weekend (which instead represent the lowest values of mobility). Figure 16 also shows that the decrease is greater when restrictions in the form of risk zones are most severe (*i.e.* when many regions are under red zone restrictions, see also Figure 1).

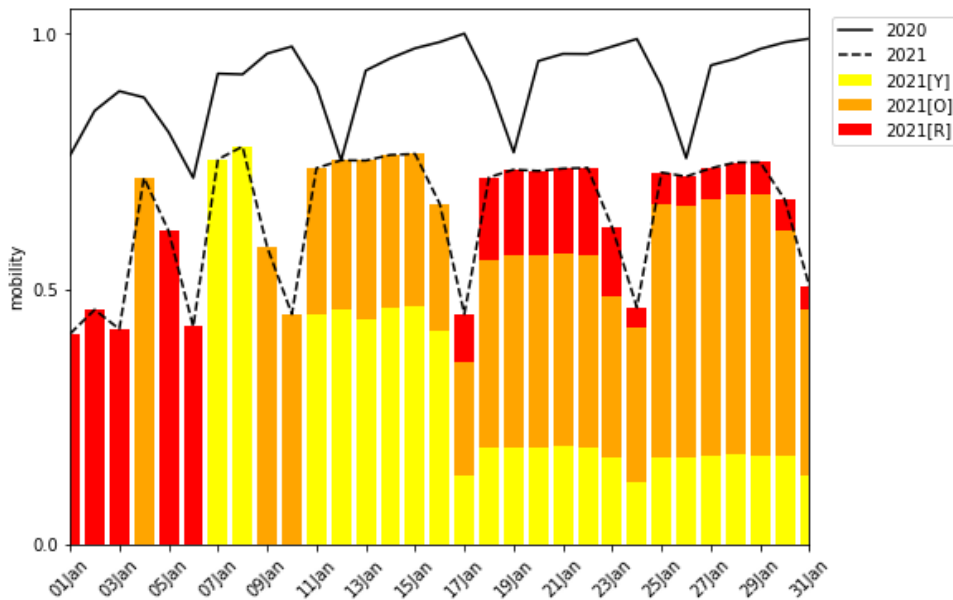

Figure 16: Comparison between total mobility in January 2020 (solid black line) versus January 2021 (dashed black line) differentiated by risk zones (see Figure 19) components (bar graph).

Table 3 quantifies the variation of mobility in January 2021 compared to the previous year for each region. A decrease in mobility is observed for all regions, with drops greater than or equal to 50% for Aosta Valley, Autonomous Province of Trento and Autonomous Province of Bolzano. The average percentage reduction of mobility per region is -29.9%. The distribution of mobility variation does not show a clear correlation with the number of days of high, medium or low restrictions. As an example, the lowest set of mobility restrictions (5 days of *red*, 3 days of *orange* and 23 days of *yellow* zones) applied to distinct regions resulted in different mobility variations: -53% in the Autonomous Province of Trento and -17.3% in Molise.

| Region | Variation % Jan 2020 - Jan 2021 | Red days | Orange days | Yellow days | Nr cases per 100k inhabitants |
|---|---|---|---|---|---|
| Aosta Valley | -57.0% | 5 | 18 | 8 | 6236 |
| Aut. Prov. Trento | -53.2% | 5 | 3 | 23 | 5047 |
| Aut. Prov. Bolzano | -50.0% | 20 | 3 | 8 | 7541 |
| Veneto | -41.6% | 5 | 24 | 2 | 6398 |
| Friuli Venezia Giulia | -39.3% | 5 | 18 | 8 | 5599 |
| Emilia-Romagna | -35.4% | 5 | 24 | 2 | 4898 |
| Lombardy | -34.6% | 12 | 17 | 2 | 5366 |
| Piedmont | -32.4% | 5 | 18 | 8 | 5175 |
| Marche | -28.5% | 5 | 18 | 8 | 3668 |
| Liguria | -27.8% | 5 | 18 | 8 | 4569 |
| Apulia | -26.0% | 5 | 18 | 8 | 3109 |
| Sicily | -25.5% | 20 | 9 | 2 | 2792 |
| Lazio | -24.6% | 5 | 18 | 8 | 3569 |
| Tuscany | -22.6% | 5 | 3 | 23 | 3641 |
| Basilicata | -21.2% | 5 | 3 | 23 | 2391 |
| Abruzzo | -20.9% | 5 | 18 | 8 | 3311 |
| Umbria | -20.3% | 5 | 18 | 8 | 4143 |
| Campania | -20.2% | 5 | 3 | 23 | 3890 |
| Molise | -17.3% | 5 | 3 | 23 | 2780 |
| Sardinia | -16.1% | 5 | 11 | 15 | 2391 |
| Calabria | -14.4% | 5 | 24 | 2 | 1729 |

Table 3: January 2021 percentage mobility variation compared to January 2020 for each Italian region, with number of days of *red*, *orange*, and *yellow* zones in January 2021, and number of cumulative COVID-19 cases per 100.000 inhabitants as of 31 January 2021.

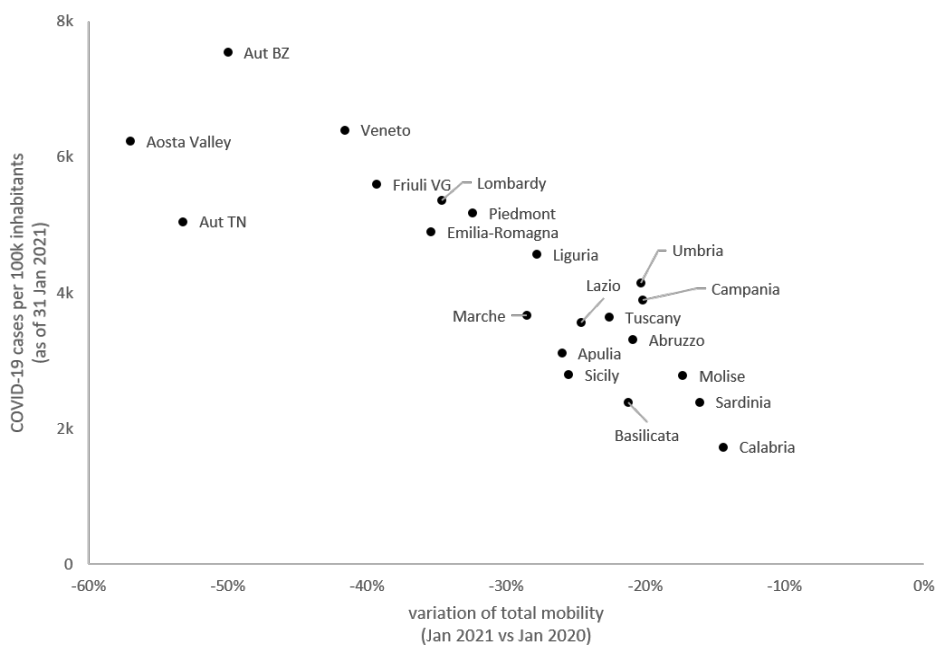

Figure 17: Relationship between mobility variation between January 2021 and January 2020 in Italian regions and the total number of COVID-19 cases as of 31 January 2021.

Nevertheless, as shown in Figure 17, the variation of mobility reduction is correlated to the number of COVID-19 cases in the different regions provided by the Italian Department of Civil Protection[13] and normalised to the resident population data provided by ISTAT[14]. This suggests a variable impact of the measures depending on the incidence of the virus, greater where the number of total cases since the start of the pandemic has been greater.

## 5  Economic impact analysis

Despite being a relative sudden phenomenon, the dramatic impact of the COVID-19 pandemic on the world has triggered numerous research efforts aimed at estimating its economic impact (see e.g. (van der Wielen and Barrios, 2020, Maliszewska et al., 2020, Sarkodie and Owusu, 2020, Makridis and Hartley, 2020, McKibbin and Fernando, 2020, Atkeson, 2020, Maital and Barzani, 2020)). We contribute to this literature by proposing an admittedly simple but effective method to assess the economic impact of mobility restrictions, via a nowcasting approach. In order to do so, a simplified analysis of economic impact based on the relationships between demand of electricity and GDP and electricity and mobility is performed. GDP monthly estimates are available until August 2020 at World Bank while energy demand in GWh is available until December 2020 through the monthly report by Terna[15]. As the *mobility* term, the monthly average of the *total* mobility indicators ad the national scale has been used. All values are normalised to their respective levels levels on January 2020. Table 4 shows the correlation between the three dimensions.

|          | gdp  | energy | mobility |
|----------|------|--------|----------|
| gdp      | 1    | 0.81   | 0.76     |
| energy   | 0.81 | 1      | 0.90     |
| mobility | 0.76 | 0.90   | 1        |

Table 4: Spearman correlation coefficient among mobility, energy demand and GDP.

As in (Chen et al., 2020), the relationship between GDP and electricity demand at monthly frequency and at country level is estimated with the following simple regression:

$$\text{GDP}_t = \alpha_0 + \alpha_1 \text{energy}_t + \alpha_2 \text{energy}^2 + \epsilon_t, \quad t = \text{Jan19}, \ldots, \text{Aug20}. \tag{1}$$

The quadratic term seems to captures better the relationship between the two variables. The results of the regression are reported in Table 5

A similar equation for the demand of electricity and mobility, also at national level, is then estimated

$$\text{electricity}_t = \beta_0 + \beta_1 \text{mobility}_t, \quad t = \text{Jan20}, \ldots, \text{Dec20}. \tag{2}$$

In this case, the additional quadratic term does not give any contribution, hence a simple linear relationship is adopted. The results are reported in Table 6.

At this point, model Eq. (1) allows to forecast the GDP indicators for September, October, November and December 2020, and the combination of Eq. (2) and again Eq. (1) allows to nowcast also January 2021, although it might be noticed that the January 2021 nowcast is subject to the forecast error in energy demand due to model Eq. (2) times the forecast error of model Eq. (1). The results are reported in Table 7 with confidence intervals[16].

One could think to use directly the mobility to describe GDP but, in fact, it is not yet known which portion of mobility reduction corresponds to teleworking, therefore it is assumed that the demand of electricity is a better proxy of productivity. Also, on the technical side, the linear and nonlinear models for GDP vs mobility does not, in fact, fit as well as the model based on energy consumption.

---

[13]Italian Department of Civil Protection, COVID-19 data repository: https://github.com/pcm-dpc/COVID-19
[14]Italian National Institute of Statistics - ISTAT, Resident population on 1st January by region: http://dati.istat.it/Index.aspx?DataSetCode=DCIS_POPRES1
[15]Terna Rete Italia is the company that deals with the national electricity grid's operation, maintenance and development in Italy.
[16]Prediction intervals are 1 percentage point larger on average.

|                     | *Dependent variable:* |
|---------------------|:---------------------:|
|                     | GDP                   |
| energy              | 3.652**               |
|                     | (1.441)               |
|                     |                       |
| energy^2            | −0.017**              |
|                     | (0.008)               |
|                     |                       |
| constant            | −92.089               |
|                     | (67.248)              |
| Observations        | 20                    |
| $R^2$               | 0.531                 |
| Adjusted $R^2$      | 0.476                 |
| Residual Std. Error | 4.275 (df = 17)       |
| F Statistic         | 9.638*** (df = 2; 17) |
| *Note:*             | *p<0.1; **p<0.05; ***p<0.01 |

Table 5: Results of the estimation of the regression model Eq. (1)

|                     | *Dependent variable:* |
|---------------------|:---------------------:|
|                     | energy                |
| mobility            | 0.395***              |
|                     | (0.097)               |
|                     |                       |
| constant            | 57.911***             |
|                     | (8.525)               |
| Observations        | 12                    |
| $R^2$               | 0.622                 |
| Adjusted $R^2$      | 0.585                 |
| Residual Std. Error | 5.587 (df = 10)       |
| F Statistic         | 16.476*** (df = 1; 10) |
| *Note:*             | *p<0.1; **p<0.05; ***p<0.01 |

Table 6: Results of the estimation of the regression model Eq. (2)

| **Month**  | **2019** | **2020**       | **2021**       |
|------------|----------|----------------|----------------|
| January    | 100.8    | 100.0          | **93.9** (±2.8)|
| February   | 100.8    | 99.9           |                |
| March      | 100.7    | 86.2           |                |
| April      | 100.7    | 85.4           |                |
| May        | 100.7    | 84.9           |                |
| June       | 100.7    | 88.5           |                |
| July       | 100.6    | 92.1           |                |
| August     | 100.5    | 95.7           |                |
| September  | 100.4    | **99.0** (±2.3)|                |
| October    | 100.3    | **98.6** (±2.3)|                |
| November   | 100.2    | **97.6** (±2.4)|                |
| December   | 100.1    | **98.3** (±2.3)|                |

Table 7: Italian Gross Domestic Product, in relative terms with respect to January 2020: the values in bold are nowcasting estimates. In parentheses the 95% confidence interval limits of estimates.

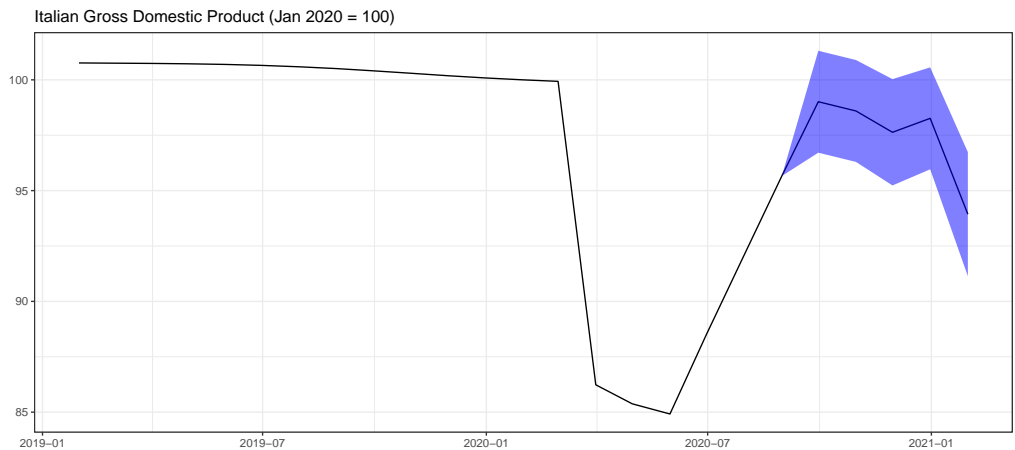

Figure 18: Plot of Italian Gross Domestic Product, in relative terms with respect to January 2020: The blue ribbon represents the 95% confidence interval limits of the nowcast estimates.

## 5.1 Regional impact on GDP

As seen in Table 3 there are quite relevant regional differences in mobility reduction given the same policy measures as it is also well known that each region contribute differently to the national GDP of the country. In order to estimate the change in GDP gain/loss during the different periods the steps are:

— to estimate the daily energy demand using daily total mobility for region $r$ and model Eq. (2);

— to estimate the GDP daily variation at regional level via model Eq. (1);

— for each transition from, e.g., *yellow → orange*, *orange → red*, etc, to estimate the average loss/gain of GDP by region and then weighting accordingly to the relative contribution of each region to the national GDP (see Table 8).

| Region | Contribution | Region | Contribution |
|---|---|---|---|
| Abruzzo | 1.92% | Liguria | 2.83% |
| Aosta Valley | 0.28% | Lombardy | 22.00% |
| Apulia | 4.35% | Marche | 2.45% |
| Aut. Prov. Bolzano | 1.41% | Molise | 0.37% |
| Aut. Prov. Trento | 1.16% | Piedmont | 7.78% |
| Basilicata | 0.71% | Sardinia | 1.98% |
| Calabria | 1.89% | Sicily | 5.06% |
| Campania | 6.11% | Tuscany | 6.67% |
| Emilia-Romagna | 9.15% | Umbria | 1.27% |
| Friuli Venezia Giulia | 2.16% | Veneto | 9.26% |
| Lazio | 11.21% | | |

Table 8: Relative contribution of each region to the national GDP for year 2018. The relative contribution were calculated using NUTS2-level Gross Domestic Product data provided by Eurostat.

This way, the bias due to regions that contribute less in terms of GDP but also experience less mobility reduction or vice-versa is avoided. The results of the analysis are given in Table 9 in which the average daily GDP variation compared to previous period is reported. Remark that these are not absolute losses/gain compared to a *business as usual* situation, but only with respect to previous stringency measure status, i.e., there is no transition from *white* regime to yellow, for example. Further, these are daily effects and have to be considered cumulatively for the length of the period in each stringency status to get an overall loss/gain of GDP at national level. We do not go into these details, but we notice that the effects are consistent with expectations: *i)* the stronger the stringency measures, the higher the GDP loss and *ii)* that there is symmetry in these variations, therefore the length of each status is a key element to determine the final balance of stringency measures on GDP.

| Transition | GDP average daily % variation |
|---|---|
| yellow → orange | -0.15% |
| yellow → red | -0.33% |
| orange → red | -0.22% |
| red → orange | +0.24% |
| orange → yellow | +0.16% |

Table 9: Average daily % variation of Gross Domestic Product compared to the previous status, depending on the transitions.

# 6 Conclusions

The present work examines the impact on the mobility and the economy of the restrictive measures adopted to fight COVID-19 in Italy. The study focuses on the period from 6 November 2020 to 31 January 2021, during which a three-tier system based on different levels of risk was adopted and applied at regional level during the second wave of the pandemic. Mobile Network Operator anonymised and aggregate mobility data shared in the context of a Business-to-Government initiative with the European Commission were used to quantify the impact of these restrictions on the mobility. Furthermore, mobility data variations are used as a proxy to estimate the impact of the restrictions on the economy in different areas of the country.

The analyses were performed at different spatial resolution, from the national and regional up to the city level. The results indicate that the impact observed for larger scale (*i.e.* regional), also applies to the smaller ones. For example, the drop of total mobility at NUTS2 (regional) level is in line with the reduction of mobility at NUTS3 (province) level.

Overall, each restriction (from low-*yellow*, to high-*red*), has proved effective in systematically reducing the mobility. In particular, transitioning to high-risk *red* zones resulted in a reduction of total movements of -28% (-26.2% outward movements) and -36.9% (-40.5% outward movements) when downgrading from *orange* and *yellow* zones respectively. When comparing mobility in January 2021 and January 2020, *i.e.* during and before mobility restrictions respectively, the results show a mobility reduction of -29.9% on average. However, this varies substantially between regions. Despite a limited correlation between the variation of mobility and the number of days of high, medium or low restrictions, the variation of mobility seems correlated to the number of COVID-19 cases in the different regions, suggesting an impact of the restrictions on mobility in direct proportion to the epidemiological situation. In other words, with the same level of restriction defined by the three-tier system, the resulting mobility reduction varies across regions and correlates with the epidemiological situation.

In general, once a region transits into a zone with further restrictions, after an initial drop, a systematic slight increase of mobility is observed over time. This relaxation tendency to the restrictions, can be quantified to an approximate 7% per week while the level of restrictions remain the same.

A significant impact of the three-tier system on Italian economic activity has been measured via a nowcasting framework applied on the gross domestic product (GDP), based on the relationships between demand of electricity and GDP, and electricity and mobility. The overall variations of GDP depending on each type of transition are coherent with the strengthening/weakening of the restrictions, ranging from an average daily variation of -0.33% for a *yellow* to *red* transition, to +0.24% for a *red* to *orange* one). These variations have to be taken with several caveats: they are daily average variations, and so, to obtain an overall loss/gain of GDP at a national level they need to be considered cumulatively for the length of the period in which the mobility restriction measure is taken and, more importantly, they represent a rough estimation, worth of further (and more refined) mathematical and statistical modelling.

This analysis can be further complemented and extended to other countries covered by the Business-to-Government data sharing initiative between European Mobile Network Operators and the European Commission.

Finally, the results can be used by national and local authorities to better design and implement targeted policies and measures that balance between the socio-economic impact of the mobility restriction and their epidemiological outcome.

Annexes

Figure 19 shows the three-tier system based on different levels of risk set up by the Italian government, with the relative restrictions.

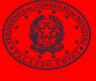

Figure 19: Italian three-tier system based on different levels of risks, in force since November 2020, source: Italian Government.

Table 6 shows the main legislative provisions of the Italian government to contain the spreading of the virus and manage the epidemiological emergency between November 2020 and January 2021.

| Publi-cation | Legislative provision | Entry into force | Link | Notes |
|---|---|---|---|---|
| 4 Nov 2020 | DECRETO DEL PRESIDENTE DEL CONSIGLIO DEI MINISTRI 3 novembre 2020. - Ulteriori disposizioni attuative del decreto-legge 25 marzo 2020, n. 19, convertito, con modificazioni, dalla legge 25 maggio 2020, n. 35, recante 'Misure urgenti per fronteggiare l'emergenza epidemiologica da COVID-19', e del decreto-legge 16 maggio 2020, n. 33, convertito, con modificazioni, dalla legge 14 luglio 2020, n. 74, recante 'Ulteriori misure urgenti per fronteggiare l'emergenza epidemiologica da COVID-19'. (20A06109) (GU Serie Generale n.275 del 04-11-2020 - Suppl. Ordinario n. 41) | 6 Nov 2020 | https://www.gazzettaufficiale.it/eli/id/2020/11/04/20A06109/sg | Definition of the three-tier system based on different levels of risk. General curfew between 22:00 and 05:00. |
| 5 Nov 2020 | MINISTERO DELLA SALUTE. ORDINANZA 4 novembre 2020 - Ulteriori misure urgenti in materia di contenimento e gestione dell'emergenza epidemiologica da COVID-19. (20A06144) (GU Serie Generale n.276 del 05-11-2020) | 6 Nov 2020 | https://www.gazzettaufficiale.it/eli/id/2020/11/05/20A06144/sg | Association of the three-tier system based on different levels of risk to all regions. |
| 10 Nov 2020 | MINISTERO DELLA SALUTE. ORDINANZA 10 novembre 2020 - Ulteriori misure urgenti in materia di contenimento e gestione dell'emergenza epidemiologica da COVID-19. (20A06211) (GU Serie Generale n.280 del 10-11-2020) | | https://www.gazzettaufficiale.it/eli/id/2020/11/10/20A06211/sg | Association of the three-tier system based on different levels of risk to all regions. |
| 14 Nov 2020 | MINISTERO DELLA SALUTE. ORDINANZA 13 novembre 2020 - Ulteriori misure urgenti in materia di contenimento e gestione dell'emergenza epidemiologica da COVID-19. (20A06292) (GU Serie Generale n.284 del 14-11-2020) | | https://www.gazzettaufficiale.it/eli/id/2020/11/14/20A06292/sg | Association of the three-tier system based on different levels of risk to all regions. |
| 20 Nov 2020 | MINISTERO DELLA SALUTE. ORDINANZA 19 novembre 2020 - Ulteriori misure urgenti in materia di contenimento e gestione dell'emergenza epidemiologica da COVID-19. (20A06423) (GU Serie Generale n.289 del 20-11-2020) | | https://www.gazzettaufficiale.it/eli/id/2020/11/20/20A06423/sg | Misure urgenti contenimento contagio nelle regioni: Calabria, Lombardia, Piemonte, Puglia, Sicilia, Valle d'Aosta. |
| 21 Nov 2020 | MINISTERO DELLA SALUTE. ORDINANZA 20 novembre 2020 - Ulteriori misure urgenti in materia di contenimento e gestione dell'emergenza epidemiologica da COVID-19. (20A06467) (GU Serie Generale n.290 del 21-11-2020) | | https://www.gazzettaufficiale.it/eli/id/2020/11/21/20A06467/sg | Misure urgenti contenimento contagio nella regione Abruzzo. |
| 23 Nov 2020 | MINISTERO DELLA SALUTE. ORDINANZA 24 novembre 2020 - Ulteriori misure urgenti in materia di contenimento e gestione dell'emergenza epidemiologica da COVID-19. (20A06541) (GU Serie Generale n.292 del 24-11-2020) | | https://www.gazzettaufficiale.it/eli/id/2020/11/24/20A06541/sg | Misure urgenti contenimento contagio nelle regioni: Basilicata, Liguria, Umbria e della Provincia autonoma di Bolzano. |
| 28 Nov 2020 | MINISTERO DELLA SALUTE. ORDINANZA 27 novembre 2020 - Ulteriori misure urgenti in materia di contenimento e gestione dell'emergenza epidemiologica da COVID-19. (20A06657) (GU Serie Generale n.296 del 28-11-2020) | | https://www.gazzettaufficiale.it/eli/id/2020/11/28/20A06657/sg | Misure urgenti contenimento contagio nelle regioni: Campania, Emilia-Romagna, Friuli-Venezia Giulia, Marche e Toscana. |
| 2 Dec 2020 | DECRETO-LEGGE 2 dicembre 2020, n. 158. Disposizioni urgenti per fronteggiare i rischi sanitari connessi alla diffusione del virus COVID-19. (20G00184) (GU Serie Generale n.299 del 02-12-2020) | 3 Dec 2020 | https://www.gazzettaufficiale.it/eli/id/2020/12/02/20G00184/sg | Extension from 30 to 50 days of 3 Nov 2020 decree. National red zone and ban to travel between regions on 25, 26 Dec 2020. Special provision for all regions from 21 Dec 2020 to 6 Jan 2021. |
| 3 Dec 2020 | LEGGE 27 novembre 2020, n. 159. Conversione in legge, con modificazioni, del decreto-legge 7 ottobre 2020, n. 125, recante misure urgenti connesse con la proroga della dichiarazione dello stato di emergenza epidemiologica da COVID-19 e per la continuita' operativa del sistema di allerta COVID, nonche' per l'attuazione della direttiva (UE) 2020/739 del 3 giugno 2020. (20G00182) (GU Serie Generale n.300 del 03-12-2020) | 4 Dec 2020 | https://www.gazzettaufficiale.it/eli/id/2020/12/03/20G00182/sg | |
| 12 Dec 2020 | MINISTERO DELLA SALUTE. ORDINANZA 11 dicembre 2020 - Ulteriori misure urgenti in materia di contenimento e gestione dell'emergenza epidemiologica da COVID-19. Modifica della classificazione delle Regioni Abruzzo, Basilicata, Calabria, Lombardia e Piemonte. (20A06975) (GU Serie Generale n.308 del 12-12-2020) | | https://www.gazzettaufficiale.it/eli/id/2020/12/12/20A06975/sg | New risk level for regions: Abruzzo, Basilicata, Calabria, Lombardia e Piemonte. |
| 18 Dec 2020 | DECRETO-LEGGE 18 dicembre 2020, n. 172. Ulteriori disposizioni urgenti per fronteggiare i rischi sanitari connessi alla diffusione del virus COVID-19. (20G00196) (GU Serie Generale n.313 del 18-12-2020). Note: Decreto-Legge convertito con modificazioni dalla L. 29 gennaio 2021, n. 6 (in G.U. 30/01/2021, n. 24) | 19 Dec 2020 | https://www.gazzettaufficiale.it/eli/id/2020/12/18/20G00196/sg | Urgent measures for Christmas holidays and the beginning of the new year. |
| 19 Dec 2020 | MINISTERO DELLA SALUTE. ORDINANZA 18 dicembre 2020 - Ulteriori limitazioni agli ingressi nel territorio nazionale. (20A07121) (GU Serie Generale n.314 del 19-12-2020) | | https://www.gazzettaufficiale.it/eli/id/2020/12/19/20A07121/sg | |
| 20 Dec 2020 | MINISTERO DELLA SALUTE. ORDINANZA 20 dicembre 2020 - Ulteriori misure urgenti in materia di contenimento e gestione dell'emergenza epidemiologica da COVID-19. (20A07122) (GU Serie Generale n.315 del 20-12-2020) | | https://www.gazzettaufficiale.it/eli/id/2020/12/20/20A07122/sg | Limitation of movements from UK and Northern Ireland. |
| 23 Dec 2020 | MINISTERO DELLA SALUTE. ORDINANZA 23 dicembre 2020 - Ulteriori misure urgenti in materia di contenimento e gestione dell'emergenza epidemiologica da COVID-19. (20A07212) (GU Serie Generale n.318 del 23-12-2020) | | https://www.gazzettaufficiale.it/eli/id/2020/12/23/20A07212/sg | Limitation of movements from UK and Northern Ireland. |
| 5 Jan 2021 | DECRETO-LEGGE 5 gennaio 2021, n. 1 (Raccolta 2021). Ulteriori disposizioni urgenti in materia di contenimento e prevenzione dell'emergenza epidemiologica da COVID-19. (21G00001) (GU Serie Generale n.3 del 05-01-2021) | 6 Jan 2021 | https://www.gazzettaufficiale.it/eli/id/2021/01/05/21G00001/sg | Ban to travel between regions from 7 to 15 Jan 2021. Possibility to move between municipalities with less than 5000 inhabitants. |
| 11 Jan 2021 | MINISTERO DELLA SALUTE. ORDINANZA 9 gennaio 2021 - Ulteriori misure urgenti in materia di contenimento e gestione dell'emergenza epidemiologica da COVID-19. (21A00137) (GU Serie Generale n.7 del 11-01-2021) | | https://www.gazzettaufficiale.it/eli/id/2021/01/11/21A00137/sg | Limitation of movements from UK and Northern Ireland. |
| 14 Jan 2021 | DECRETO-LEGGE 14 gennaio 2021, n. 2 - Ulteriori disposizioni urgenti in materia di contenimento e prevenzione dell'emergenza epidemiologica da COVID-19 e di svolgimento delle elezioni per l'anno 2021. (21G00002) (GU Serie Generale n.10 del 14-01-2021) | 14 Jan 2021 | https://www.gazzettaufficiale.it/eli/id/2021/01/14/21G00002/sg | Introduction of a fourth level of risk (White zone). Others. |
| 15 Jan 2021 | DECRETO DEL PRESIDENTE DEL CONSIGLIO DEI MINISTRI 14 gennaio 2021. - Ulteriori disposizioni attuative del decreto-legge 25 marzo 2020, n. 19, convertito, con modificazioni, dalla legge 22 maggio 2020, n. 35, recante 'Misure urgenti per fronteggiare l'emergenza epidemiologica da COVID-19', del decreto-legge 16 maggio 2020, n. 33, convertito, con modificazioni, dalla legge 14 luglio 2020, n. 74, recante 'Ulteriori misure urgenti per fronteggiare l'emergenza epidemiologica da COVID-19', e del decreto-legge 14 gennaio 2021 n. 2, recante 'Ulteriori disposizioni urgenti in materia di contenimento e prevenzione dell'emergenza epidemiologica da COVID-19 e di svolgimento delle elezioni per l'anno 2021'. (21A00221) (GU Serie Generale n.11 del 15-01-2021 - Suppl. Ordinario n. 2) | | https://www.gazzettaufficiale.it/eli/id/2021/01/15/21A00221/sg | |
| 18 Jan 2021 | MINISTERO DELLA SALUTE. ORDINANZA 16 gennaio 2021 - Ulteriori misure urgenti in materia di contenimento e gestione dell'emergenza epidemiologica da COVID-19. (21A00237) (GU Serie Generale n.13 del 18-01-2021) | | https://www.gazzettaufficiale.it/eli/id/2021/01/18/21A00237/sg | Limitation of movements from Brazil. |
| 23 Jan 2021 | MINISTERO DELLA SALUTE. ORDINANZA 22 gennaio 2021 - Ulteriori misure urgenti in materia di contenimento e gestione dell'emergenza epidemiologica da COVID-19 per le Regioni Calabria, Emilia Romagna e Veneto. (21A00402) (GU Serie Generale n.18 del 23-01-2021) | 24 Jan 2021 | https://www.gazzettaufficiale.it/eli/gu/2021/01/23/18/sg/html | As of 24 Jan: Calabria, Emilia Romagna and Veneto stay orange. |
| 23 Jan 2021 | MINISTERO DELLA SALUTE. ORDINANZA 22 gennaio 2021 - Ulteriori misure urgenti in materia di contenimento e gestione dell'emergenza epidemiologica da COVID-19 per la Regione Sardegna. (21A00401) (GU Serie Generale n.18 del 23-01-2021) | 23 Jan 2021 | https://www.gazzettaufficiale.it/eli/gu/2021/01/23/18/sg/html | As of 23 Jan: Sardinia from yellow to orange. |
| 23 Jan 2021 | MINISTERO DELLA SALUTE. ORDINANZA 23 gennaio 2021 - Ulteriori misure urgenti in materia di contenimento e gestione dell'emergenza epidemiologica da COVID-19 per la Regione Lombardia. (21°00403) (GU Serie Generale n.18 del 23-01-2021) | 24 Jan 2021 | https://www.gazzettaufficiale.it/eli/gu/2021/01/23/18/sg/html | As of 24 Jan: Lombardy from red to orange. |
| 29 Jan 2021 | MINISTERO DELLA SALUTE. ORDINANZA 29 gennaio 2021 - Ulteriori misure urgenti in materia di contenimento e gestione dell'emergenza epidemiologica da COVID-19 per le Regioni Puglia, Sicilia, Umbria e per la Provincia autonoma di Bolzano. (21A00537) (GU Serie Generale n.25 del 31-01-2021) | 1 Feb 2021 | https://www.gazzettaufficiale.it/atto/serie_generale/caricaDettaglioAtto/originario?atto.dataPubblicazioneGazzetta=2021-01-31&atto.codiceRedazionale=21A00537 | As of 1 Feb: Puglia and Umbria stay in orange; Sicilia and Provincia autonoma di Bolzano switch from red to orange. |
| 29 Jan 2021 | MINISTERO DELLA SALUTE. ORDINANZA 29 gennaio 2021 - Ulteriori misure urgenti in materia di contenimento e gestione dell'emergenza epidemiologica da COVID-19 per le Regioni Calabria, Emilia Romagna, Lombardia e Veneto. (21A00536) (GU Serie Generale n.25 del 31-01-2021) | 1 Feb 2021 | https://www.gazzettaufficiale.it/eli/id/2021/01/31/21A00536/sg | As of 1 Feb: Calabria, Emilia Romagna, Lombardia e Veneto switch from orange to yellow. |
| 30 Jan 2021 | MINISTERO DELLA SALUTE. ORDINANZA 30 gennaio 2021 - Ulteriori misure urgenti in materia di contenimento e gestione dell'emergenza epidemiologica da COVID-19. (21A00535) (GU Serie Generale n.24 del 30-01-2021) | 1 Feb 2021 | https://www.gazzettaufficiale.it/eli/id/2021/01/30/21A00535/sg | The three-tier system based on different levels of risk is extended to 15 Feb. |

Table 11: Legislative provisions of the Italian government to contain the spreading of COVID-19 pandemic and to manage the epidemiological emergency from November 2020 - January 2021.

KJ-NA-30630-EN-N

### The European Commission's science and knowledge service
Joint Research Centre

### JRC Mission
As the science and knowledge service of the European Commission, the Joint Research Centre's mission is to support EU policies with independent evidence throughout the whole policy cycle.

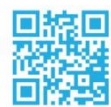
**EU Science Hub**
ec.europa.eu/jrc

@EU_ScienceHub

EU Science Hub – Joint Research Centre

Joint Research Centre

EU Science Hub

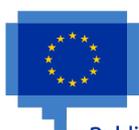
Publications Office
of the European Union

doi:10.2760/241286
ISBN 978-92-76-32080-7